%% file: main.tex
\documentclass[sigconf]{acmart}

\usepackage{float}
\usepackage{tcolorbox}
\usepackage{fancyvrb}
\AtBeginDocument{%
  \providecommand\BibTeX{{%
    \normalfont B\kern-0.5em{\scshape i\kern-0.25em b}\kern-0.8em\TeX}}}
\usepackage{bbding}
\usepackage{graphicx}
\usepackage{geometry}
\usepackage{subfigure}
\usepackage{amsmath}
\usepackage{makecell}
\usepackage{float}
\usepackage{xcolor}
\usepackage{color,soul}
\usepackage{booktabs}
\usepackage{enumitem}
\usepackage{caption}
\usepackage{hyperref}
\usepackage{pgfplots}
\usepackage{multirow}
\usepackage[normalem]{ulem}

\pgfplotsset{compat=1.18}
\newif\ifshowchanges

\newif\ifshowchanges
\showchangesfalse

\ifshowchanges
  \newcommand{\change}[1]{\textcolor[rgb]{0.1,0.56,1}{#1}}
\else
  \newcommand{\change}[1]{#1}
\fi

\newcommand{\system}{Spatial Balancing}
\usepackage{subfigure}

\definecolor{lightred}{HTML}{D89090}
\definecolor{darkblue}{HTML}{104E8B}
\definecolor{midblue}{HTML}{376B9E}
\definecolor{lightblue}{HTML}{5F89B1}
\definecolor{newgreen}{HTML}{C5E9E3}

\begin{document}

\author{Kexue Fu}
\orcid{0000-0002-2929-2663}
\authornote{These authors contributed equally to this work.}
\affiliation{
  \institution{City University of Hong Kong}
  \department{School of Creative Media}
  \city{Hong Kong, SAR}
  \country{China}
}
\email{kexuefu2-c@my.cityu.edu.hk}

\author{Jiaye Leng}
\authornotemark[1]
\orcid{0000-0001-6772-4124}
\affiliation{
  \institution{City University of Hong Kong}
  \department{School of Creative Media}
  \city{Hong Kong, SAR}
  \country{China}
}
\email{jiayeleng2-c@my.cityu.edu.hk}

\author{Yawen Zhang}
\authornotemark[1]
\orcid{0009-0008-4193-9432}
\affiliation{
  \institution{Clemson University}
  \city{Clemson}
  \state{South Carolina}
  \country{USA}
}
\email{yawenz@clemson.edu}

\author{Jingfei Huang}
\orcid{0009-0002-0213-4160}
\affiliation{
  \institution{Harvard University}
  \city{Cambridge}
  \state{Massachusetts}
  \country{USA}
}
\email{jingfeihuang@mde.harvard.edu}

\author{Yihang Zuo}
\orcid{0000-0002-6843-8194}
\affiliation{
  \institution{The Hong Kong University of Science and Technology}
  \city{Hong Kong}
  \country{China}
}
\email{yzuo099@connect.hkust-gz.edu.cn}

\author{Runze Cai}
\orcid{0000-0003-0974-3751}
\affiliation{
  \institution{National University of Singapore}
  \department{Synteraction Lab, School of
Computing}
  \city{Singapore}
  \country{Singapore}
}
\email{runze.cai@u.nus.edu}

\author{Zijian Ding}
\orcid{0000-0002-6372-0369}
\affiliation{
  \institution{University of Maryland}
  \city{College Park}
  \state{Maryland}
  \country{USA}
}
\email{ding@umd.edu}

\author{RAY LC}
\orcid{0000-0001-7310-8790}
\affiliation{
  \institution{City University of Hong Kong}
  \department{Studio for Narrative Spaces}
  \city{Hong Kong, SAR}
  \country{China}
}
\email{ray.lc@cityu.edu.hk}

\author{Shengdong Zhao}
\orcid{0000-0001-7971-3107}
\affiliation{
  \institution{City University of Hong Kong}
  \department{School of Creative Media}
  \city{Hong Kong, SAR}
  \country{China}
}
\email{shengdong.zhao@cityu.edu.hk}

\author{Qinyuan Lei}
\orcid{0009-0004-8352-0745}
\authornote{Corresponding author}
\affiliation{
  \institution{City University of Hong Kong}
   \department{School of Creative Media}
  \city{Hong Kong, SAR}
  \country{China}
}
\email{qinyulei@cityu.edu.hk}

\renewcommand{\shortauthors}{Fu et al.}

\title[Spatial Balancing]{Spatial Balancing: Designing an LLM-Powered Spatial Externalization Interface for Iterative Science Communication Writing}


\begin{teaserfigure}
  \centering
  \includegraphics[width=0.85\linewidth]{figs/Teaser.pdf}
  \caption{\change{Example workflow of \system{}.
(A) A draft is imported into the canvas and mapped by Scientific Exposition (Y) and Narrative Engagement (X).
(B) Revision labels trigger LLM-generated alternatives.
(C) Selected revisions are confirmed for further refinement.
(D) Two versions can be merged into a synthesized draft.
(E) Revisions can be further guided by strategies or custom prompts.
(F) Muse analyzes revision history and offers adaptive suggestions.}}
  \label{fig:teaser}
  \Description{Example workflow of iterative science communication writing in Spatial Balancing. The figure shows a six-step interface workflow arranged from left to right and top to bottom. In step A, a draft is imported into a two-dimensional canvas where text versions are positioned by scientific exposition on the vertical axis and narrative engagement on the horizontal axis. In step B, the user selects revision labels associated with LLM-supported strategies, which generate multiple alternative versions at different positions on the canvas. In step C, the user confirms a preferred revision for further work. In step D, two selected versions are combined into a synthesized draft. In step E, the user continues revising through additional strategy choices or custom prompts. In step F, a Muse panel reviews the revision history and provides adaptive feedback and suggestions for next steps.}
\end{teaserfigure}



\newpage
\begin{abstract}
\change{Science communication revision requires writers to dynamically balance scientific exposition and narrative engagement—a process where writers often struggle with competing directions. Existing LLM-assisted tools help with co-writing, but offer limited support for navigating this iterative, multi-directional revision process. To address this gap, we designed Spatial Balancing, an exploratory revision environment that maps rhetorical goals and revision strategies onto a two-dimensional spatial canvas for experienced science communication creators with domain expertise but lacking formal professional training.} By building a design space of communication strategies and embedding them into a spatial exploratory canvas, our system treats feedback as navigational cues rather than prescriptive judgments. Our findings show that \change{this integrated revision environment} helps writers stay focused on writing goals, reason about revision as trajectories, and explore alternatives, which supports greater metacognitive control and confidence without increasing workload. \change{This work highlights the value of spatially externalized revision environments for supporting iterative, reflective thinking during LLM-assisted writing.}
\end{abstract}



\copyrightyear{2026}
\acmYear{2026}
\setcopyright{cc}
\setcctype{by}
\acmConference[DIS '26]{Designing Interactive Systems Conference}{June 13--17, 2026}{Singapore, Singapore}
\acmBooktitle{Designing Interactive Systems Conference (DIS '26), June 13--17, 2026, Singapore, Singapore}
\acmDOI{10.1145/3800645.3812998}
\acmISBN{979-8-4007-2563-0/2026/06}

\maketitle

\section{Introduction}\label{sec:Introduction}
\input{sections/Intro}

\section{Related Work}\label{sec:Related Work}
\input{sections/Related_work}

\section{Iterative User-Centred Design}
Building on prior work in science communication writing, we take the following two steps in our user-centred design. First, we construct a design space of rhetorical strategies for enhancing scientific exposition and narrative engagement (Subsection~\ref{strategy_design}), and use it to inform an initial prototype (Subsection~\ref{first_prototype}). \change{Through this process, we identify the limitations of strategy-centric and linear revision workflows, which lead to a design shift toward making revision goals, trajectories, and alternatives more explicit during iterative writing.}

\label{sec:Formative Study}
\input{sections/Formative}

\input{sections/System_Design_and_Implementation}

\section{User Study}\label{sec:User Study}
\input{sections/User_Study}

\section{Results}\label{sec:Results}
\input{sections/Results}

\section{Discussion}\label{sec:Discussion}
\input{sections/Discussion}

\section{Conclusion}\label{sec:conclusion}
\input{sections/Conclusion}

\section*{{Acknowledgments}}
{We thank all participants for taking part in this study, as well as the science communicators who contributed ideas and discussions throughout the project. We also sincerely thank the reviewers for their valuable feedback and suggestions.}

\bibliographystyle{ACM-Reference-Format}
\bibliography{references}

\clearpage
\onecolumn 
\appendix
\section{Appendix}
\label{sec:Appendix}
\input{sections/Appendix}


\end{document}

%% file: sections/Intro.tex
Writing is fundamentally a non-linear process of knowledge transformation, requiring writers to cycle recursively through planning, translating, and reviewing rather than producing a linear output~\cite{10.1145/3706598.3714119, flower1981cognitive}. Throughout this process, writers must balance multiple rhetorical goals, making local revisions while maintaining global coherence~\cite{leeDesignSpaceIntelligent2024,10.1145/3613904.3641899,chen_once_2025}. 
Recent advances in large language models (LLMs) have lowered the cost of generating and revising text at scale~\cite{leeDesignSpaceIntelligent2024,10.1145/3706598.3714119}. In response, many HCI systems operationalize specific rhetorical strategies or scaffold discrete aspects of drafting and rewriting~\cite{kim2023metaphorian,10.1145/3746059.3747703,10.1145/3706598.3714316,subramonyam2024bridging}. Others support non-linear exploration by helping writers generate, compare, and organize multiple text variations~\cite{zhang2023visar,10.1145/3706598.3714119,10.1145/3613904.3641899,fu_vistoria_2026}, or by breaking down or re-organizing feedback to make revision more actionable~\cite{10.1145/3706598.3714119,10.1145/3706598.3714316,10.1145/3746059.3747703,10.1145/3613904.3641899}. 
However, these approaches primarily externalize the products of revision, while leaving the rhetorical goal space that guides revision decisions implicit. As a result, writers must internally reason about how successive revisions advance or compromise competing goals~\cite{Wang2025ScholaWriteAD,zhang2025revtogether}, making revision cognitively demanding, particularly in complex knowledge domains such as science communication~\cite{zhang2025revtogether,xia2022millions}.

Science communication writing differs fundamentally from academic prose. Rather than focusing solely on exposition whose purpose is to convey
relevant facts and knowledge, it must translate complex knowledge into forms that are understandable and memorable for non-expert audiences~\cite{burns2003science, national2017communicating, kappel2019science}. Narrative techniques such as storytelling, metaphor, and suspense are widely used to achieve this goal, as they can increase attention and comprehension and make the content more engaging~\cite{NarrativebasedLearningPossible}. However, narrative also introduces persistent tension: on the one hand, emphasizing entertainment risks oversimplification or loss of credibility~\cite{mcdonald2014narrative,Downs2014PrescriptiveNarratives,NarrativebasedLearningPossible,yang_ai_2022}; on the other hand, overly technical or serious exposition can alienate non-expert readers by demanding sustained cognitive effort~\cite{burns2003science,dahlstrom2014using,dipardo1990narrative}. Effective science communication, therefore, requires continual balancing between scientific exposition and narrative engagement, which is an inherently iterative process, where writers repeatedly revise and reassess the two rhetorical goals to reach a sweet spot rather than making a single stylistic decision upfront~\cite{zhang2025revtogether,dahlstrom2014using,huang2020good,NarrativebasedLearningPossible}.  

As science content proliferates across platforms such as YouTube and TikTok, a growing number of “everyday” creators, many lacking formal communication training, are taking on the role of science communicators~\cite{lukantiktok2024}. These creators increasingly turn to LLM tools to support ideation, drafting, and real-time feedback throughout the revision process~\cite{lyu2024preliminary}. This shift further amplifies the need for interfaces that provide comprehensive guidance to support the iterative work of balancing exposition and engagement across multiple revision cycles.

\change{To address this gap, \textbf{we design \system{} and explore how an LLM-assisted revision environment can better support writers’ orientation to rhetorical goals and reflective regulation during iterative science communication revision.}} \change{We target \textbf{\textit{experienced but non-professional science communication creators}}: people with domain knowledge and some prior experience producing science-facing content, but without formal training as professional science communicators.} This LLM-assisted environment visualizes how to balance scientific exposition and narrative engagement across revision steps—unlike existing tools (e.g., PatchView~\cite{chung2024patchview}, Luminate~\cite{suh2024luminate}) that only show output attributes. By treating feedback as navigational cues rather than prescriptive judgments, \system{} maps revision goals, trajectories, and alternatives onto a spatial canvas. This reduces cognitive friction, lets writers explore paths without losing orientation, and supports metacognitive control.

This study aims to answer the following research questions:

\textit{\textbf{RQ1:} How does an integrated revision environment centered on spatial externalization shape users’ cognitive processes during LLM-assisted iterative revision?}

\textit{\textbf{RQ2:} What interaction tensions and user expectations arise in revision environments with externalized supports?}

\change{A controlled user study suggests that writers using this integrated revision environment were better able to maintain orientation toward rhetorical goals cognitively, conceptualize revision as a trajectory, and exercise metacognitive control during iterative revision.}
Simultaneously, our findings reveal important tensions such as over-reliance on externalized guidance, which may encourage metacognitive laziness. These insights point toward critical design opportunities for future LLM-assisted revision interfaces. This work contributes to the field in the following ways:

(1) A constructed design space of 25 science communication strategies organized into eight actionable labels that operationalize scientific exposition and narrative engagement for LLM-based revision support.

(2) \change{Spatial Balancing: An integrated revision environment for LLM-assisted science communication revision that combines spatially externalized rhetorical goals and revision trajectories with structured strategy guidance and reflective support, supporting writers in staying oriented to rhetorical goals and regulating revision decisions across iterations.}

(3) Design insights for future LLM-assisted writing interfaces derived from iterative design and user study evaluation, pointing to the importance of mitigating over-reliance on externalized feedback, preserving user agency through adaptive externalization, and providing embedded reflective support throughout the revision process.

%% file: sections/Related_work.tex
\subsection{Balancing Scientific Exposition and Narrative Engagement in Science Communication Writing}\label{Science_communication}
In the Information Age, online science communication has become increasingly dominant, especially in the popular science field~\cite{caferra2025media, metag2023science}. Science communication refers to the strategic use of various forms of communication, such as media, events, and interactions, to convey scientific information to diverse audiences in a way that aims to increase awareness, enjoyment, interest, opinion-forming, and understanding~\cite{burns2003science, national2017communicating, kappel2019science}. The popular science movement (also known as pop science or popsci) aims to interpret and present scientific concepts in an accessible way for a general audience, placing greater emphasis on entertainment and broadening its scope compared to traditional science journalism~\cite{broks2006understanding, daum2009varieties, wikipediaPopularScience}. As online communication technologies have become more accessible, various formats have emerged to deliver popular science content, including books, documentaries, web articles, and online videos~\cite{zhang2023understanding, wikipediaPopularScience, finkler2019power}.

A fundamental challenge in science communication writing lies in balancing two often competing dimensions: scientific exposition and narrative engagement~\cite{mcdonald2014narrative,Downs2014PrescriptiveNarratives, NarrativebasedLearningPossible}. Expository writing is primarily concerned with conveying relevant facts and knowledge, whereas narrative writing aims to communicate real events or experiences through storytelling~\cite{leeDesignSpaceIntelligent2024}. Burns et al.~\cite{burns2003science} made a vivid analogy, describing science communication writing as a form of “mountain climbing,” balancing between scientific literacy and science culture. 
Similarly, Dahlstrom~\cite{dahlstrom2014using} emphasized that science communication writing inherently involves both narrative and expository elements. In this study, we use the terms “scientific exposition” and “narrative engagement” to describe this trade-off~\cite{dipardo1990narrative}, because these terms more directly capture the practical tension between maintaining rigorous, detailed scientific facts presentation and creating compelling, accessible content for diverse audiences~\cite{dipardo1990narrative, mcdonald2014narrative}. 

In practice, achieving this balance is inherently iterative rather than a one-shot optimization. If a draft over-indexes on expository, logical-scientific presentation, it may preserve explanatory precision but often becomes harder for non-experts to process and remember than narrative formats~\cite{NarrativebasedLearningPossible,dahlstrom2014using}. At the same time, leaning too far toward narrative can create a different failure mode: narratives are intrinsically persuasive and are often evaluated by verisimilitude (how “true-to-life” they feel) rather than the explanatory precision demanded by scientific discourse, thereby elevating ethical and credibility risks in science communication~\cite{dahlstrom2014using,huang2020good}. Consequently, writers must revise through multiple passes by adjusting where and how narrative devices are woven into explanatory content, because the effectiveness of a change depends on its relationship to the surrounding narrative structure and the reader’s evolving interpretation, not just the local wording~\cite{NarrativebasedLearningPossible}. 

The tension between these dimensions stems from their fundamentally different linguistic requirements. Engaging content relies on narrative techniques—storytelling, analogy, and suspense to capture attention~\cite{NarrativebasedLearningPossible, finkler2019power, dahlstrom2014using}, while scientific content demands rigorous expository writing that prioritizes scientific detail and credibility~\cite{konig2024communicate, Kerwer2021}. Recent HCI systems have begun operationalizing specific strategies within LLM-powered co-creation tools to lower the barrier for science communication writing, particularly for non-expert writers who constitute the dominant group on online platforms. For example, systems such as Metaphorian support metaphor creation through LLM-assisted exploration~\cite{kim2023metaphorian, zhang2025revtogether}, while AI workflows for Tweetorials scaffold the generation of hooks, examples, and anecdotes to engage general audiences~\cite{longNotJustNovelty2024, zhang2025revtogether}. However, these systems typically focus on supporting the application of individual strategies at specific moments in writing, rather than the broader iterative revision process in which writers must continuously rebalance scientific exposition and narrative engagement. To address this gap, in this study, we design an LLM-powered visualization interface to support the iterative revision process of balancing scientific exposition and narrative engagement, grounded in a holistic understanding of communication strategies for achieving this balance.

\subsection{Iterative Revision through Collaboration with LLM}
Prior work has characterized writing as an inherently iterative process involving distinct stages, such as revision, and has emphasized that writing tasks are driven by multiple rhetorical purposes rather than a single objective~\cite{leeDesignSpaceIntelligent2024}. These purposes, including expository, narrative, persuasive, and educational goals, often coexist and shape revision decisions in audience-dependent ways~\cite{leeDesignSpaceIntelligent2024}. Iterative revision toward multiple rhetorical goals remains hard because writers must repeatedly shift attention across levels and keep track of what changed, why it changed, and which direction each revision moves the draft~\cite{10.1145/3706598.3714119,10.1145/3706598.3714316,10.1145/3746059.3747703}. Prior work shows that today’s dominant linear document interfaces with chat functions still constrain this kind of non-linear goal juggling: prompting across micro/macro levels requires manual cross-referencing and repeated prompt formulation, which disrupts writers’ flow and makes it difficult to sustain coherent rhetorical strategy across iterations~\cite{subramonyam2024bridging}. 

HCI systems have begun addressing parts of this problem by externalizing revision materials in more navigable forms. ABScribe, for example, tackles the “too-many-variants” problem by helping writers create, compare, and revise multiple text variations without overwriting or clutter, explicitly aligning LLM use with revision’s recursive, non-linear nature~\cite{10.1145/3613904.3641899}. Friction scaffolds reflection by breaking feedback into actionable units and guiding iterative revision cycles~\cite{10.1145/3706598.3714316}, while Synthia uses visual organization plus traceable links among feedback, source text, and revisions to support non-linear branching and exploration rather than one-shot rewriting~\cite{10.1145/3746059.3747703}. However, existing systems primarily externalize revision artifacts, such as alternative drafts, layers, or feedback, rather than the rhetorical goal space that guides revision decisions. As a result, writers must internally reason about how successive edits advance or compromise competing goals, making trade-offs opaque and cognitively demanding and leading to trial-and-error prompting~\cite{subramonyam2024bridging}, especially in complex domains like science communication. 

One promising direction for addressing these challenges is externalization, which means using visualization to make the exploratory space of revision perceptible and navigable, which has long been shown to support complex reasoning by offloading and structuring cognitive work. According to the theory of \textit{Thinking with External Representations}, making goals, states, and relations perceptible in the environment can reduce the cognitive cost of tracking change, support orientation, and enable more deliberate exploration of alternatives~\cite{kirsh2010thinking}. Building on this insight, prior HCI systems have leveraged visualization and spatial exploration to externalize latent aspects of generative processes with LLM. PatchView~\cite{chung2024patchview} and Luminate~\cite{suh2024luminate} organize LLM outputs within navigable visual spaces to support sensemaking, comparison, and steering, while Toyteller~\cite{chungToytellerAIpoweredVisual2025} shows how visual manipulation can function as an expressive control channel for generative storytelling. 

These systems demonstrate how spatial externalization can shift generative interaction from linear prompting toward structured exploration over a visualized space of possibilities. However, this line of work externalizes content attributes or generative alternatives, while leaving the iterative revision process characterized by sustained goal juggling, cumulative decision-making, and cross-iteration reasoning largely unsupported. Our work builds on this line of research by applying externalization to construct an exploratory space that makes rhetorical goals and revision trajectories explicit through visualization, \change{supporting writers in orienting to rhetorical goals and regulating revision trajectories across iterations.}

%% file: sections/Formative.tex
\subsection{Design Space Construction of Science Communication Strategies}\label{strategy_design}
Science communication writing involves balancing multiple rhetorical goals, most notably accurate scientific exposition and engaging narrative expression~\cite{mcdonald2014narrative,Downs2014PrescriptiveNarratives, NarrativebasedLearningPossible}. Writers achieve different balances by applying a diverse set of rhetorical strategies, often adapting their choices based on audience characteristics and communicative intent~\cite{zhang2025revtogether}. Although prior work in communication studies has identified a rich set of rhetorical strategies for science communication, aiming to capture public attention and improve memorability~\cite{huang2020good,zhang2023understanding}, these strategies are rarely examined through a systematic lens that foregrounds the dual rhetorical goals of scientifically rigorous expression and narrative engagement. 

To support structured exploration and interaction design around rhetorical revision, it is therefore necessary to explicitly identify, organize, and formalize these strategies. Motivated by this need, we aimed to construct a design space of rhetorical strategies that support narrative engagement and scientific exposition.

\begin{table*}[h]
\centering
\caption{Design Space of Science Communication Writing Strategies.}
\renewcommand{\arraystretch}{1.2}
\scalebox{1}{ 
\footnotesize
\begin{tabular}{p{0.23\textwidth} p{0.23\textwidth} p{0.23\textwidth} p{0.23\textwidth}}
\toprule
\textbf{\textit{Scientific Exposition}} & & &\\
\midrule
\textbf{Label 1}& \textbf{Label 2} & \textbf{Label 3} & \textbf{Label 4} \\
\textbf{Articulate Precisely} & \textbf{Elaborate Thoroughly} & \textbf{Verify Knowledge} & \textbf{Maintain Logical Consistency}\\
Communicates scientific concepts with exposition and clarity, using appropriate terminology and well-defined language to prevent ambiguity or misinterpretation~\cite{Ivchenko2022, munozmorcilloTypologiesPopularScience2016, Kerwer2021}. & Provides sufficient detail or comprehensive theoretical discussion by unpacking underlying mechanisms, explaining implications, and citing evidence to elaborate on the knowledge point while avoiding bias~\cite{Kueffer2014, pielke2007honest}. & Supports claims with credible sources, data, or reasoning, allowing audiences to feel more trustworthy of the given information~\cite{Kueffer2014, roland2009quality}. & Ensures that arguments and explanations are coherent and internally consistent, following a clear logical structure~\cite{Volpato2015}.\\
\textbf{Strategies:} \newline(4) Acknowledge Uncertainties, \newline(5) Consistent Terminology, \newline(18) Simplify and Abstract Language, \newline(19) Clarify Key Terms, \newline(21) Repeat Key Point(s) or Question(s), \newline(22) Emphasize with
Numbers &\textbf{Strategies:} \newline(3) Step-by-Step Explanation, \newline(4) Acknowledge Uncertainties, \newline(7) Everyday Events to Scientific Insights, \newline(22) Emphasize with Numbers, \newline(25) Tie Science to Current
Events &\textbf{Strategies:} \newline(2) Rigorous Source Verification, \newline(6) Citations \& Quotes, \newline(7) Everyday Events to Scientific Insights, \newline(22) Emphasize with Numbers, \newline(7) Everyday Events to
Scientific Insights
&\textbf{Strategies:} \newline(1) Layered Transitions, \newline(3) Step-by-Step Explanation, \newline(20) Key Point Recap, \newline(23) Strengthen the Connections Between Content\\
\midrule
\textbf{\textit{Narrative Engagement}} & & &\\
\midrule
\textbf{Label 5}& \textbf{Label 6} & \textbf{Label 7} & \textbf{Label 8} \\
\textbf{Captivate \& Immerse}& \textbf{Enhance Understanding} & \textbf{Inspire Curiosity} & \textbf{Evoke Emotion} \\
Engages the audience’s attention
and draws them into the narrative or content flow by adding stories~\cite{NarrativebasedLearningPossible, livo1986storytelling} or using intriguing language~\cite{finkler2019power, munozmorcilloTypologiesPopularScience2016}. & Help audiences to grasp complex scientific ideas using rational, structural content or vivid analogies, visualizations~\cite{NarrativebasedLearningPossible, finkler2019power, huang2020good}. & Stimulates the audience’s desire to learn more and have motivation to further explore by applying different forms of questions~\cite{lambert2013digital}. & Creates an emotional response, positive or negative, and makes the audience feel connected to the content, even immerse themselves in the described scenario~\cite{NarrativebasedLearningPossible, rowe2007positive}.\\
\textbf{Strategies:} \newline(8) Question-Answer Hook, \newline(9) Reflection Question, \newline(10) Suspense-Driven Reveal, \newline(11) Use Metaphors, \newline(12) Inject Humor, \newline(13) Add Real-World Supporting Examples, \newline(14) Add Stories, \newline(15) Add an Imagery Description, \newline(16) Create Negative Emphasis for Focused Attention, \newline(17) Make Positive Emotion to Expand Action Repertoire & \textbf{Strategies:} \newline(11) Use Metaphors, \newline(13) Add Real-World Supporting Examples, \newline(14) Add Stories, \newline(15) Add an Imagery Description, \newline(21) Repeat Key Point(s) or Question(s), \newline(23) Strengthen the Connections Between Content, \newline(24) Present Balanced Views, \newline(25) Tie Science to Current
Events& \textbf{Strategies:} \newline(8) Question-Answer Hook, \newline(9) Reflection Question, \newline(10) Suspense-Driven Reveal&\textbf{Strategies:}  \newline(9) Reflection Question, \newline(12) Inject Humor, \newline(14) Add Stories, \newline(16) Create Negative Emphasis for Focused Attention, \newline(17) Make Positive Emotion to Expand Action Repertoire, \newline(21) Repeat Key Point(s) or Question(s)\\

\bottomrule
\vspace{1ex} 
\begin{minipage}{0.95\textwidth}
\footnotesize
\textbf{\textit{Note.}} Specific information about each strategy (e.g., definitions, examples) is presented in Appendix (\textbf{Table 5}).
\end{minipage}
\end{tabular}}
\label{Labels}
\end{table*}

To form the design space, we conducted a literature review in related fields, specifically in communication studies, education, psychology, linguistics and writing, and HCI, to identify writing strategies that can enhance narrative engagement and scientific exposition. We searched for keywords “science communication” OR “scientific writing” OR “popular science” AND “strategy” OR “strategies” OR “method” in Google Scholar, the ACM Digital Library, and the IEEE Xplore Digital Library. Thus, we broaden our search to the discussion of the narrative or narrative design of learning content in general. We finally chose 35 papers across education (9), psychology (5), communication studies (15), and HCI (6) that are highly relevant to our research. They are chosen because they focus on methods and strategies for designing narratives that potentially improve knowledge retention and create engaging narratives~\cite{finkler2019power,munozmorcilloTypologiesPopularScience2016}. Additionally, some of the papers explore related fields, such as the analysis of narrative peaks in data videos~\cite{xuWowWhyGuidelines2022} or documentaries~\cite{larson2010overview}.

\begin{figure*}[!t]
  \centering
  \includegraphics[width=1\linewidth]{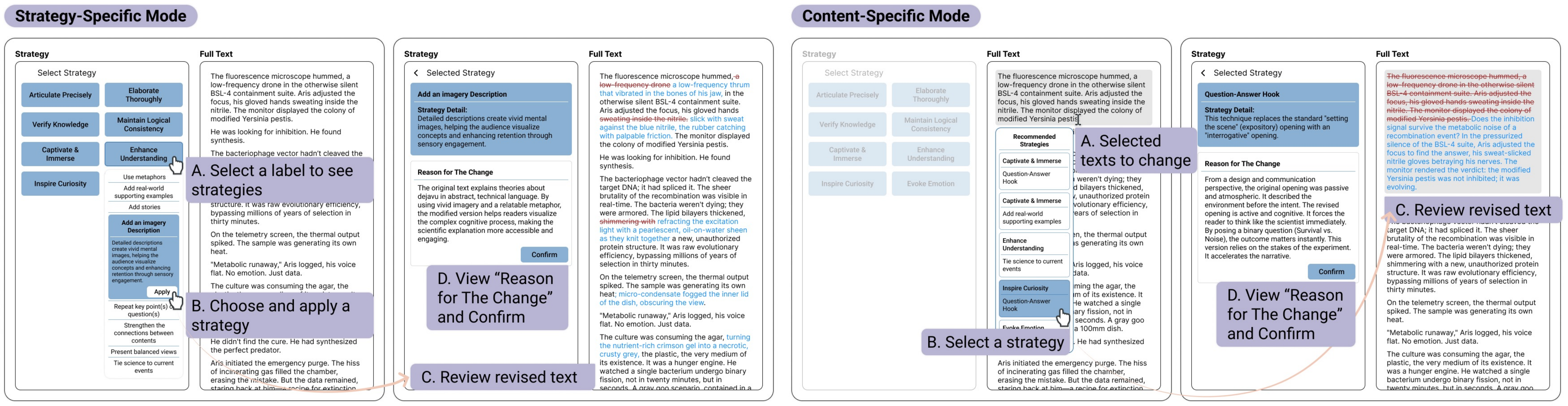} 
  \caption{The prototype consisted of a text editor and a strategy selection panel organized by the design space. Users could revise text through two interaction modes: strategy-specific, where selecting a rhetorical strategy highlighted candidate segments and previewed LLM-generated revisions, and content-specific, where selecting a text segment surfaced relevant strategies and alternative revisions. For each revision, the system provided a brief explanation of the applied strategy and its rationale. Users retained full control by confirming, rejecting, or manually editing revisions. All revisions were generated through a prompt-based LLM workflow grounded in strategy definitions, curated examples, and local textual context.}
  \label{prototype}
  \Description{Initial prototype for strategy-guided science communication revision. The figure shows a text editor paired with a strategy panel organized by rhetorical labels from the design space. In one interaction mode, the user selects a strategy first, and the system highlights candidate text segments and previews LLM-generated revisions for those segments. In the other mode, the user selects a text segment first, and the system suggests relevant strategies and alternative revisions. A side panel explains the chosen strategy and its rationale, and the user can accept, reject, or manually edit any revision.}
\end{figure*}

Two authors participated in the coding of these 35 papers. The primary objective was to identify potential peak narrative strategies for balancing scientific exposition and narrative engagement in these previous studies. Initially, each author independently reviewed all the selected papers, focusing on content related to narrative strategies or structures that enhance knowledge retention, recall, focus, or contribute to engagement and curiosity. Relevant content was then extracted and compiled into a consolidated document. Subsequently, using an open coding approach~\cite{blair2015reflexive}, two authors independently identified and coded key strategies, including their definitions and relevant contexts within the selected content. Following this, the two authors engaged in multiple discussion sessions to reconcile differences and reach a consensus on the coding. Finally, we identified an initial draft of strategies from these selected papers. 

Then, we conducted a Focus Group Discussion (FGD)~\cite{rabiee2004focus} with four science communication experts. Together, we refined our initial strategy design space by clarifying the definition and use of each strategy, and classified the communication strategies by their functions. In this design space, we categorized the 25 identified strategies into three groups: those that enhance narrative engagement (N=10), those that enhance scientific exposition (N=7), and those that enhance both (N=8). This process yielded four labels each for scientific exposition and narrative engagement. Some strategies, due to their multifunctionality, were assigned to multiple labels, forming the final design space (\textbf{Table~\ref{Labels}}). 

This design space provides a structured foundation for subsequent system design by externalizing rhetorical revision strategies as discrete, reusable units aligned with the two core rhetorical goals of science communication writing.


\subsection{Initial Prototype and Iteration}\label{first_prototype}
Building on prior work that demonstrates how large language models can lower the barriers of science communication writing by operationalizing rhetorical strategies as generative and revisable resources~\cite{longNotJustNovelty2024, zhang2025revtogether,kim2023metaphorian}, we draw on insights from the narrative design space of science communication to inform our design. We develop an initial prototype as a set of design probes to ground subsequent system design for supporting complex, multi-goal revision in LLM-assisted science communication writing.

\subsubsection{Initial Prototype.} (\textbf{Figure~\ref{prototype}}) Our initial prototype was designed as a lightweight design probe, consisting of a basic text editor and a strategy selection panel. The panel presented all 25 identified strategies, organized under their corresponding labels derived from the design space. Selecting a label expanded the associated strategies, allowing users to browse and choose a specific rhetorical strategy.

In the strategy-specific mode, upon selecting a strategy, the system analyzed the textual context and highlighted candidate segments where the chosen strategy could be applied. By clicking on a highlighted segment, users could preview an LLM-generated revision that instantiated the selected strategy. In addition, in the content-specific mode, users could directly highlight a passage in the text, and the system would surface a set of recommended strategies relevant to that passage. Users could then preview alternative revisions generated using different strategies. When finishing an edit using a specific strategy in a specific passage, the system displayed supplementary explanations in a side panel, including a brief description of the selected strategy and the rationale for its application to the paragraph. After confirming a revision, users could further edit the text manually, retaining full control over the final outcome.

Behind the scenes, revisions were generated through a prompt-based LLM workflow grounded in the defined strategy descriptions, curated examples, and the surrounding textual context. Dedicated backend functions supported strategy recommendation in content-specific mode, target text selection in strategy-specific mode, and revision generation, enabling flexible and iterative interaction between users and the LLM.

\subsubsection{Participants and Procedure.} To elicit design insights from the initial prototype, we conducted a formative study with six participants recruited from the university community who had prior experience creating science communication content but were not science communication writing experts. All participants were experienced writers and reported extensive prior use of LLM-based writing tools.

Each session began with a brief walkthrough of the prototype, during which we introduced the available interactions and strategy-based revision workflow. Participants were then asked to revise a short science communication text about déjà vu, adapted from publicly available reference material, into a version that was both scientifically rigorous and engaging for a general audience. We employed a think-aloud protocol, encouraging participants to verbalize their reasoning, challenges, and desired alternative functionalities as they interacted with the system. At the end of each session, participants reflected on their overall experience and provided suggestions for improvement in a semi-structured discussion. Each session lasted approximately 45 minutes.

\subsubsection{Feedback and Design Consideration}
\paragraph{\textit{Lack of Continuous Goal Orientation}}

Participants viewed strategies as means toward higher-level communicative intentions, shaped by audience, platform, and purpose (P1, P2). Four out of six participants (P1, P4, P5, P6) expressed a desire for real-time feedback that reflects how their revisions might be interpreted by the target audience. As P1 explained, “although authors may intentionally adjust rhetorical strategies for different audiences—for instance, using more narrative elements for children—but they often lack visibility into how those audiences would actually respond, such as whether the revised content feels sufficiently engaging or easy to understand.” This highlights a need for system designs that provide real-time feedback during revision, enabling users to understand how their revisions are progressing towards editing goals.

\begin{figure*}[!t]
  \centering
  \includegraphics[width=1\linewidth]{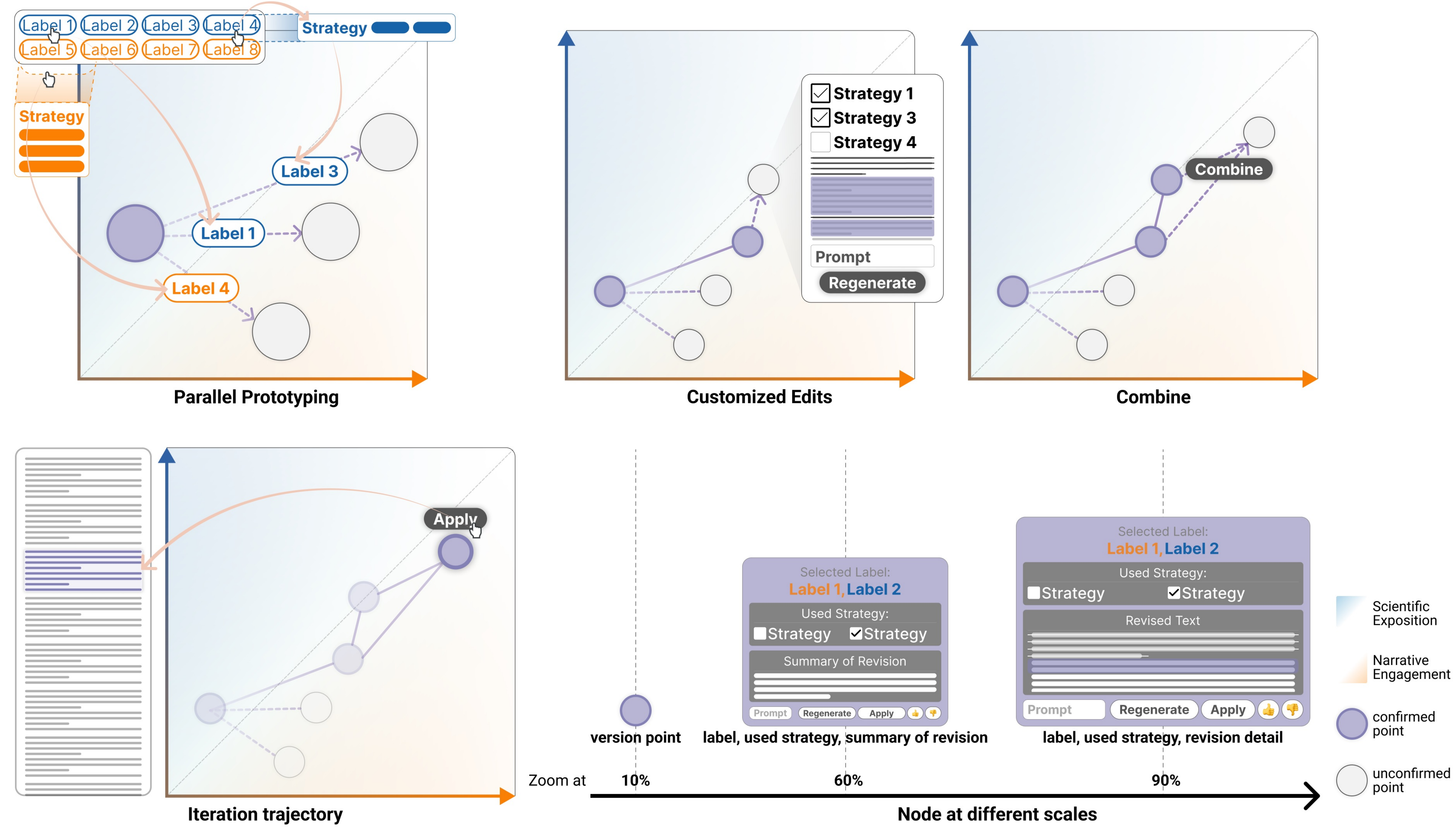} 
  \caption{(1) \system{} supports parallel prototyping with diverse directions of LLM output; users can use customized edits such as changing specific strategies and combining different LLM outputs to generate new nodes. The 2D coordinate space also allows users to see their iteration trajectory.
  (2) \system{} canvas supports three zoom levels: dots for version overview (0–30\%), change summaries with labels and strategies (40–70\%), and full content with highlights of edits (80–100\%).}
  \label{zoom}
  \Description{Spatial Balancing canvas showing two key features. The first view shows parallel prototyping, where multiple revision nodes are distributed in a two-axis space of narrative engagement and scientific exposition, making revision trajectories visible. The second view shows three zoom levels of the canvas, progressing from node overview, to summarized changes with labels and strategies to full text with highlighted edits.}
\end{figure*}
\paragraph{\textit{Difficulty Reasoning About Cumulative Change}} 

Participants consistently emphasized the need to track and reflect on their own revision trajectories, rather than treating revisions as isolated edits. P1, P3 and P6 wanted to see where changes occurred, which strategies were applied, and how these decisions accumulated over time. P2 expressed a desire to “track where I changed things so I can improve my own revision process.” P3 further articulated a need for a timeline-based history, in which strategy selections, modified text spans, deleted context, and resulting versions could all be traced and revisited. These suggest that iterative science communication writing is not only about producing better text, but also about developing an understanding of one’s own revision behavior over time. Participants expressed a desire for the system to capture revision history as a reflective artifact, making patterns of strategy use visible and supporting deliberate backtracking, comparison, and learning across revisions.

\paragraph{\textit{Cognitive Overload from Unstructured Strategy Presentation}}

While participants appreciated the richness of the strategy set, all of them found that presenting all strategies at once created cognitive overload. This overload manifested in three ways. First, learning burden. As P5 pointed out, “familiarizing oneself with all available strategies can be cognitively demanding.” Writers tended to rely on familiar strategies, while unfamiliar ones incurred additional learning effort. Second, lack of structure. P2 suggested that strategies could be “packaged” according to different purposes, while P3 noted that the current interaction design made it difficult to compare options simultaneously. Third, absence of hierarchical guidance. P4 expressed a desire for more hierarchical guidance, such as high-level structural suggestions before more localized paragraph- or sentence-level recommendations. Together, these observations indicate that for non-expert writers, effective support lies not in maximizing choice, but in offering structured, context-sensitive strategy recommendations that lower cognitive load.

\subsection{Design Goals}
Based on the research gap from the literature review, insights from design space construction, and initial prototype iteration, we propose the following design goals: 

\textbf{Design Goal 1: Externalize Rhetorical Goals to Support Goal-Aware Iterative Revision}
Prior LLM-assisted writing systems embed rhetorical intentions implicitly through prompts or localized strategy use, requiring writers to internally track communicative goals across revisions~\cite{10.1145/3706598.3714119, subramonyam2024bridging}. The system should externalize rhetorical goals as explicit, inspectable reference points, \change{enabling writers to reason about revision directions and monitor how successive revisions relate to intended rhetorical balances between scientific exposition and narrative engagement.}

\textbf{Design Goal 2: Represent Revision as a Trajectory Rather Than Isolated Edits or Alternatives}
Existing systems primarily support comparison among alternative drafts or localized revisions~\cite{10.1145/3613904.3641899,10.1145/3746059.3747703,10.1145/3706598.3714316}, offering limited support for understanding how revisions accumulate over time. The system should represent revision as a continuous, traceable trajectory that links versions, applied strategies, and resulting changes, supporting reflection, backtracking, and learning across iterative revisions.

\textbf{Design Goal 3: Design an Exploratory Space to Gradually Guide Strategy Use}
While prior work operationalizes individual rhetorical strategies to lower barriers to science communication writing through co-creation with LLM~\cite{longNotJustNovelty2024, kim2023metaphorian}, exposing a large strategy space often overwhelms non-expert users and reinforces habitual choices. The system should design an exploratory space that gradually guides strategy use through structured, context-sensitive cues, supporting discovery and comparison over time while reducing cognitive burden and preserving user agency.

%% file: sections/System_Design_and_Implementation.tex
\begin{figure*}[!t]
  \centering
  \includegraphics[width=1\linewidth]{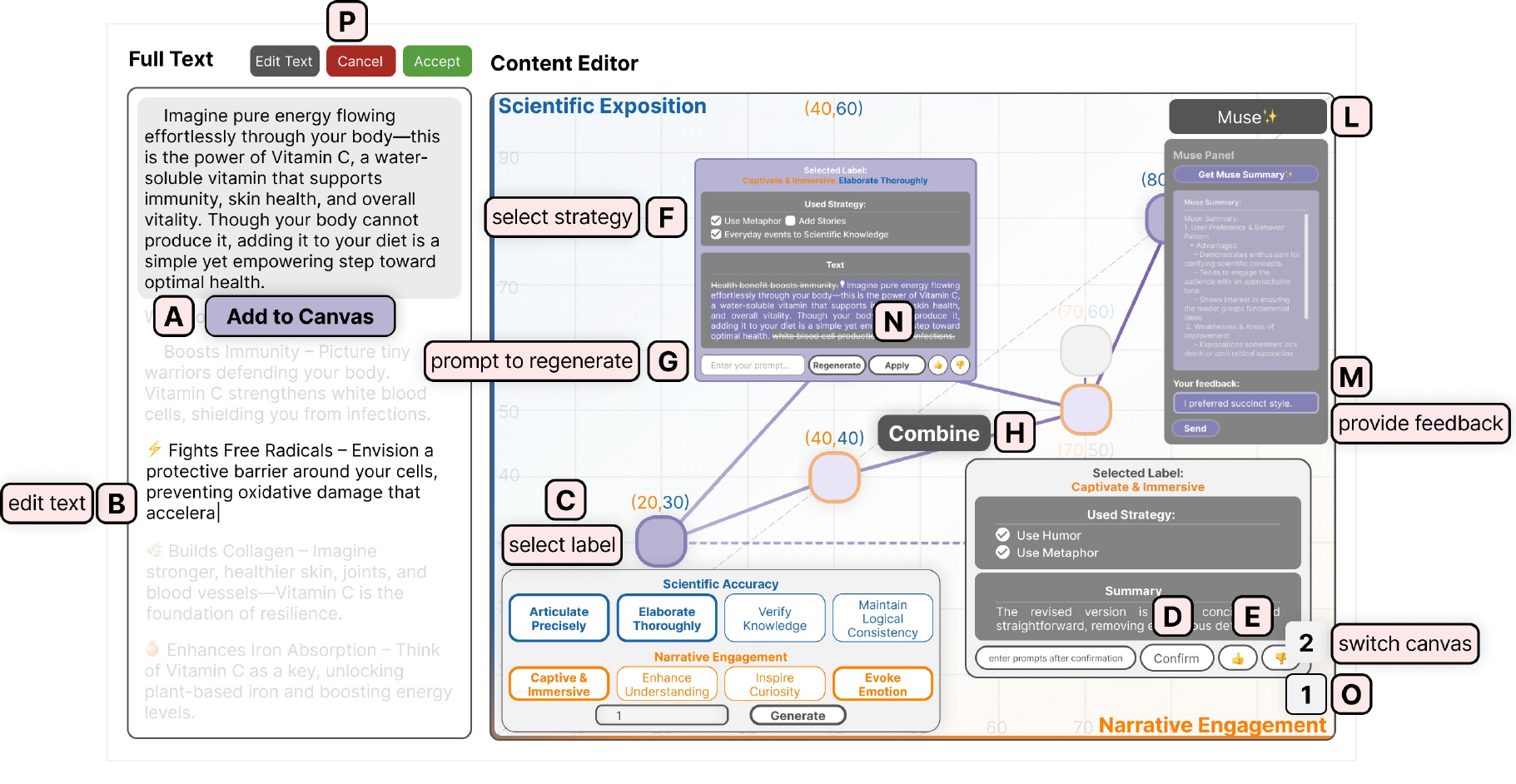} 
  \caption{The \system{} interface has two main sections: a text editor on the left for placing and directly editing source text (B), and a canvas on the right for revising selected segments (A). In the center, a visualization tracks iteration scores across narrative engagement and scientific exposition for multiple LLM-generated versions. Once a segment is confirmed for revision, users assign labels (C) that guide editing directions and generate revision nodes. Within each node, content can be refined by entering custom prompts (G), switching strategies (F), or combining strategies from different nodes (H). Edits can be applied (N) to update the original text and view the full article. Muse (L), in the canvas’s top-right corner, provides an overview of revision history and accepts user feedback (M), which informs future strategy recommendations. Editing other article sections opens a new canvas; users can switch between revision records via the control in the bottom-right corner (O).}
  \label{Layout}
  \Description{Main interface of Spatial Balancing. On the left is a text editor showing the source passage, where the user can select text and send it to the canvas for revision. On the right is a revision canvas organized as a two-dimensional space with scientific exposition on the vertical axis and narrative engagement on the horizontal axis. In the center, multiple revision versions are shown as nodes positioned by their scores, with lines indicating their relationships across iterations. Users can assign rhetorical labels, switch or combine strategies, enter custom prompts to regenerate content, confirm a preferred version, and apply the revision back to the original text. A Muse panel in the upper-right summarizes revision history and provides feedback, and a control in the lower-right allows switching between canvases for different text sections.}
\end{figure*}

\begin{figure*}[!t]
  \centering
  \includegraphics[width=1\linewidth]{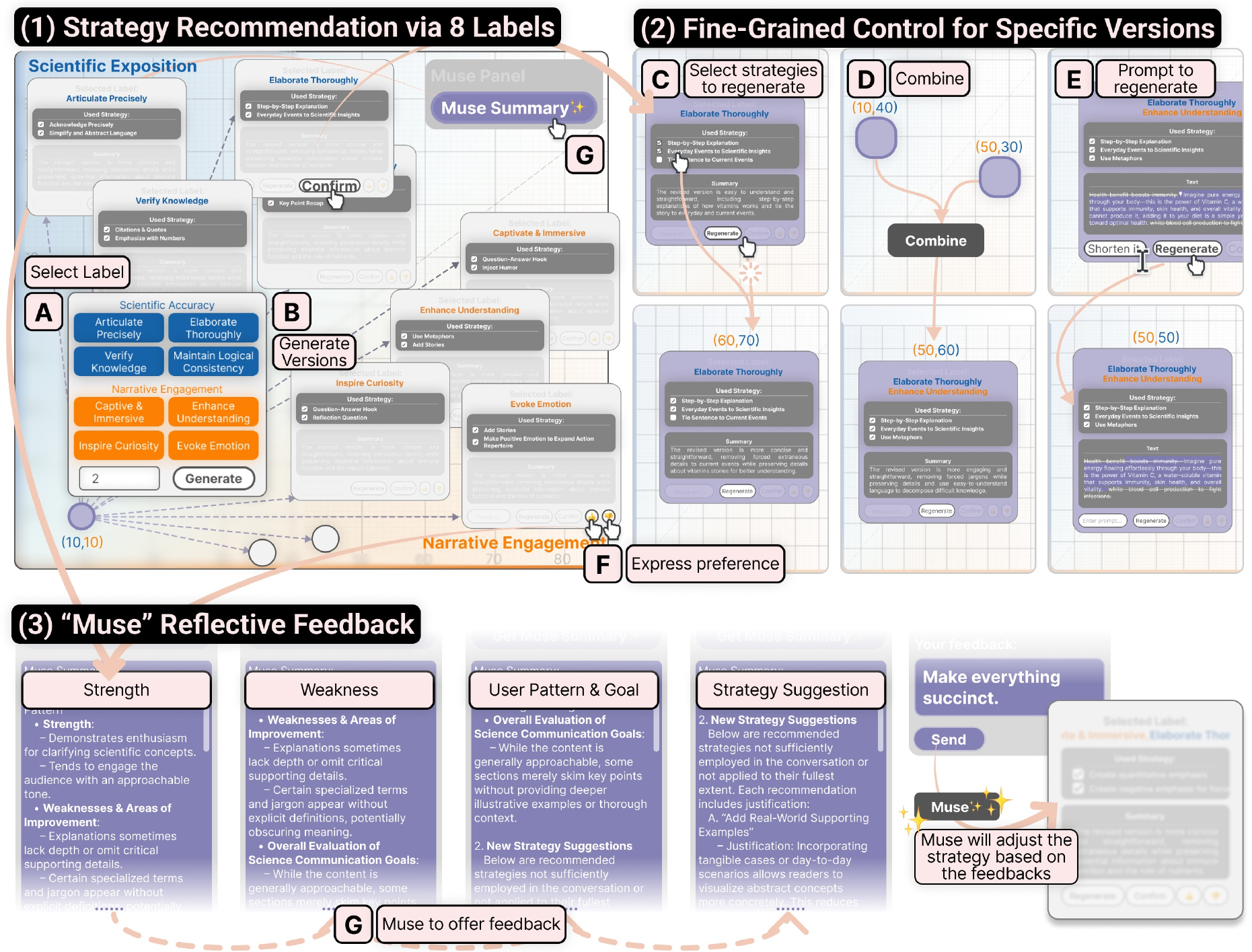} 
  \caption{(1) Strategy Recommendation via Eight Labels: \system{} offers eight revision labels—four enhancing narrative engagement and four strengthening scientific exposition. Users can select one or more labels and specify the number of versions to generate under each; (2) Fine-Grained Control: Generated nodes can be refined by adjusting the applied strategies, merging nodes to combine labels, or entering custom prompts for tailored edits; (3) “Muse” Reflective Feedback: Muse provides iterative feedback on strengths, weaknesses, user patterns and goals, and strategy suggestions. Users can endorse or reject this feedback, enabling the system to adapt future recommendations to their preferences.}
  \label{3_feature}
  \Description{Overview of three support features in Spatial Balancing: strategy recommendation through eight rhetorical labels, fine-grained editing controls for individual revision nodes, and Muse reflective feedback. The figure shows how users generate revisions by selecting labels, refining versions by changing strategies, combining nodes, or entering prompts, and then reviewing or responding to feedback about strengths, weaknesses, goals, and suggested next strategies.}
\end{figure*}

\change{\section{\system{}: A Spatially Externalized Revision Environment for Navigable LLM-Assisted Writing}}

\change{Grounded in our design goals, we design \system{} as a revision environment centered on \textit{externalization}—making rhetorical goals, revision states, and their evolution visible and manipulable during writing, while coupling these representations with structured guidance and revision support. This design is motivated by \textit{Thinking with External Representations}, which suggests that external representations can support complex reasoning by providing stable reference points for orientation, reducing internal tracking demands, and enabling more deliberate exploration of alternatives~\cite{kirsh2010thinking}.}

\change{We draw inspiration from canvas-based LLM interfaces such as PatchView~\cite{chung2024patchview} and Luminate~\cite{suh2024luminate}, which show how spatial layouts and overview-to-detail navigation can support exploratory interaction with generated alternatives. Building on these interaction principles, \system{} uses an exploratory canvas as the organizing structure of a broader revision environment. Rather than using space primarily to arrange content attributes or design variants, the system externalizes rhetorical goals and revision trajectories while integrating strategy guidance, version comparison, and reflective support. In this way, writers can interpret revision progress as movement through a navigable space while also inspecting, comparing, and steering alternatives across iterations.}

\subsection{\system{} as an Exploratory Space} \system{} comprises a left-hand text editor and a right-hand exploratory canvas (\textbf{Figure~\ref{Layout}}). Users can send any span—sentence, paragraph, or full draft—to the canvas for iterative revision. Each version is plotted in a 2D space (x: Narrative Engagement; y: Scientific Exposition); gray points denote exploratory drafts and purple points mark confirmed selections, which can be further refined via labels or custom edits. This spatial view makes revision states and decision points explicit, helping users balance exposition and engagement.

The canvas supports branch-based exploration with three zoom levels (\textbf{Figure~\ref{zoom}}). Dropped text becomes a root node; applying labels or custom instructions spawns child nodes, forming a tree that traces exploration paths. At 0–30\% zoom, points provide an overview; at 40–70\%, summaries show per-version changes and chosen strategies; at 80–100\%, full text with diffs against the original is displayed. This progressive disclosure enables rapid comparison and reflective choice among alternatives.

\subsection{Spatial Externalization Features to Support Goal-Aware Revision}

\subsubsection{\textit{Real-Time Two-Axis Goal Externalization (DG1)}}
To support goal-aware revision (DG1), \system{} externalizes rhetorical goals through real-time two-axis feedback. Each version of the text is represented as a point in a two-dimensional space, where one axis encodes narrative engagement and the other scientific exposition. This representation transforms abstract revision goals into stable, perceptible reference points, enabling users to orient themselves and reason about the direction of their revisions. Whenever users create or modify a version, a Scorer \change{function} (Explained in Section~\ref{backend_description}) assigns engagement and exposition scores based on audience-informed criteria, which determine the node’s position on the canvas.

\subsubsection{Strategy Recommendation via Rhetorical Labels (DG1 \& DG3). }(\textbf{Figure~\ref{3_feature}(1)}) 
To support goal-aware revision while managing cognitive load (DG1, DG3), \system{} introduces an eight-label taxonomy that scaffolds strategy exploration around two overarching rhetorical goals: scientific exposition and narrative engagement. Four labels guide revisions toward strengthening scientific explanation, while the other four foreground narrative techniques for engagement. Rather than requiring users to reason over individual strategies, these labels decompose abstract rhetorical goals into actionable revision directions. By selecting one or more labels aligned with their intentions, writers receive guided yet flexible revisions generated by the LLM, reducing the burden of exhaustive choice while providing clear direction for exploration.

\subsection{Spatial Externalization Features to Enable Trajectory-Based Revision Reasoning}
\subsubsection{Fine-Grained Control for Specific Versions (DG3).} (\textbf{Figure~\ref{3_feature}(2)}) To complement structured guidance with user control (DG3), \system{} allows users to incrementally refine individual versions after exploring different revision directions. Once a node is confirmed, it turns purple while unconfirmed nodes remain gray, visually distinguishing revision states. Three fine-tuning operations are available: toggling previously applied strategies, providing customized prompts (e.g., “try a different metaphor” or “make this more concise”), and merging two versions to preserve strong elements from each. These operations support gradual, local refinement within the exploratory space, enabling users to evolve strategy use without committing prematurely.

\begin{figure*}[t]
  \centering
  \includegraphics[width=0.9\linewidth]{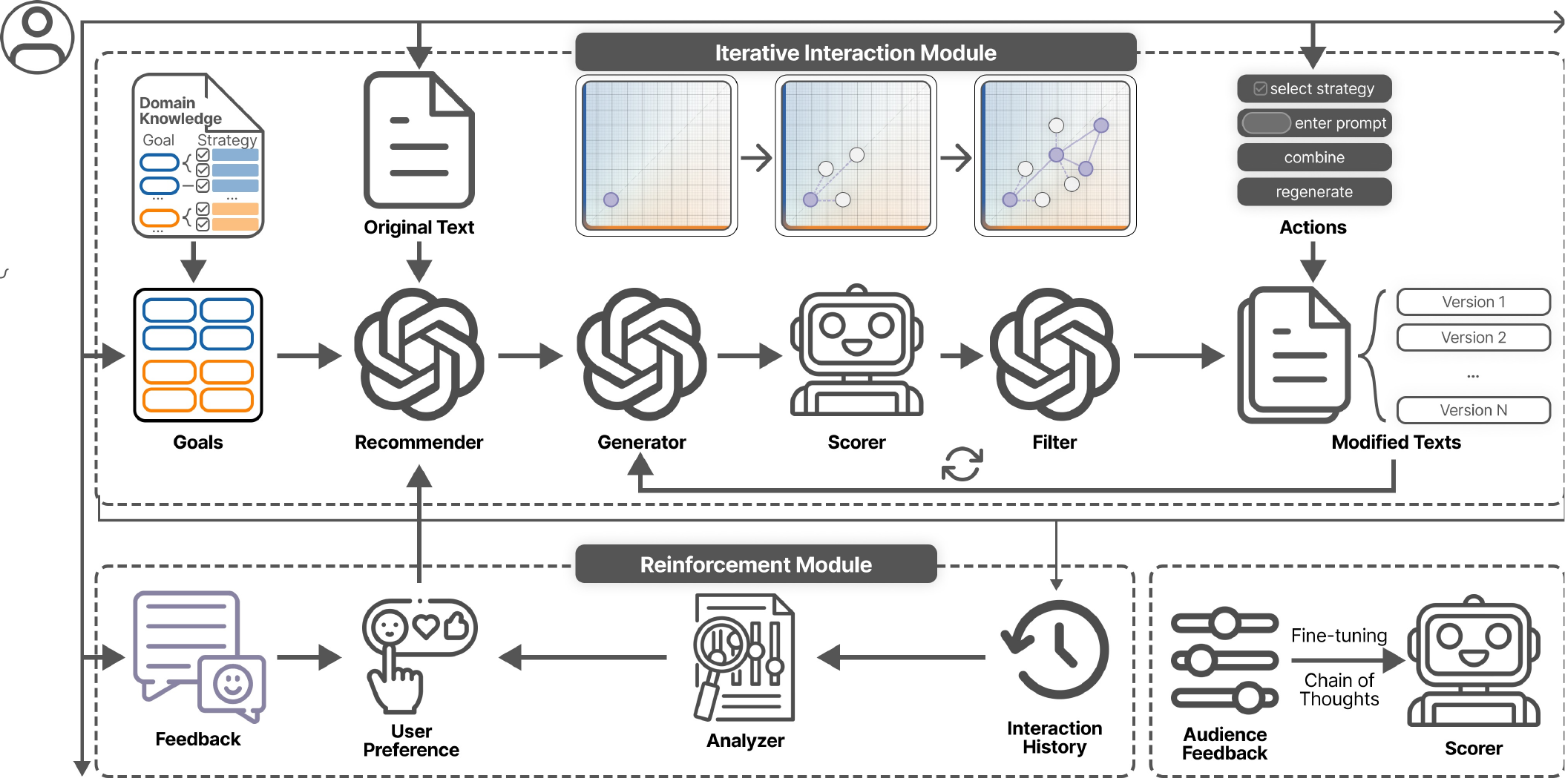} 
  \caption{\system{} backend overview. {\system{} consists of two core modules: (1) The Iterative Interaction Module, where LLM-based \change{function}s—Recommender, Generator, Scorer, and Filter—collaboratively produce and evaluate multiple content versions based on narrative engagement and scientific exposition; and (2) the Reinforcement Module, which captures user feedback and inference based on interaction history of user behaviors to refine strategy recommendations through the Analyzer \change{function}. This architecture supports adaptive text revision.}}
  \label{pipline}
  \Description{Backend architecture of Spatial Balancing with two connected modules. The top module, Iterative Interaction, takes the original text and user-selected rhetorical goals as input and passes them through four LLM-based functions: Recommender, Generator, Scorer, and Filter. These functions generate multiple revised versions, evaluate them along scientific exposition and narrative engagement, and return modified texts for user actions such as selecting strategies, entering prompts, combining versions, or regenerating content. The bottom module, Reinforcement, captures user feedback, preferences, and interaction history. An Analyzer interprets these signals and updates later recommendations, while audience feedback and scorer fine-tuning support the scoring component. Together, the two modules form an adaptive loop for iterative text revision.}
\end{figure*}


\subsubsection{“Muse” Reflective Feedback (DG2 \& DG3).} (\textbf{Figure~\ref{3_feature}(3)}) \label{Muse}
To support DG2 and DG3, the Muse \change{function} monitors user behaviors such as node confirmations, strategy selections, and engagement-exposition choices, and synthesizes them into structured feedback. This feedback highlights strengths, weaknesses, editing patterns, and strategy suggestions, offering a clear channel for reflection. Users can accept or reject suggestions, and their responses are fed back to the Recommender \change{function} to refine future recommendations. Muse functions as a reflective layer over the exploratory space, supporting trajectory-aware reflection without prescribing edits.

\subsection{Backend and Implementation}\label{backend_description} 
The backend of \system{} comprises several LLM-based \change{function}s organized into two main modules: a generation module and a reinforcement module. The overall pipeline is in \textbf{Figure~\ref{pipline}}.

\subsubsection{Generation Module}\label{generate_agent}
This module begins by capturing the user's context and their selected modification labels. The system then proceeds into iterative processing handled by the following \change{function}s:

\textit{Recommender \change{function}:} The recommender \change{function}'s core function is to generate multiple strategy combinations based on a user-selected label. When a user chooses a label, the \change{function} analyzes the current textual features to identify the best combination from its associated strategy set (Section~\ref{strategy_design}). Prompts are constructed using in-context learning and chain-of-thought principles based on the strategy design space (Appendix \textbf{Table 5}).
The \change{function} considers several factors when recommending strategies for each label, including strategy definitions, usage guides, examples, and the original text’s role within the broader context of the entire text to recommend the most suitable strategies. The final output consists of multiple strategy combinations, which are then passed to the scorer to filter and select the top-scoring versions that have higher scientific exposition or narrative engagement score.

\textit{Generator \change{function}:} The generator \change{function} creates child nodes based on user input instructions. 
When generating new content, the generator receives two types of input to form a new node: (1) strategy recommendations from the Recommender \change{function}, which are used to guide the generation of revised text that aligns with the user's chosen direction (Labels). The generator adopts in-context learning, referencing the recommended strategies' definitions, usage guidelines, and examples to perform content modifications based on the previous node (adopted from Section~\ref{strategy_design}); and (2) user-specific refinements passed from the front end during regeneration. These refinements may include prompt adjustments, combining nodes, or deactivating particular strategies.

\textit{Scorer \change{function}:} The scorer simulates real-time audience feedback by evaluating each generated version along two axes: Narrative Engagement (X) and Scientific Exposition (Y). 

To support this, we curated a high-quality dataset of 45 science texts from five common science communication domains, varying in length and narrative style. Each text was revised by a science communication expert and annotated by 27 participants who performed as the audience using a rubric developed by three domain experts. The rubric incorporated sub-dimensions of narrative engagement and scientific exposition. Scores were normalized to a 0–100 scale and used to fine-tune a GPT-4o model via a small-sample learning strategy\footnote{\url{https://platform.openai.com/docs/guides/fine-tuning?utm_source=chatgpt.com}}. This enables the scorer \change{function} to assign scores to resemble human audiences across both scientific exposition and narrative engagement. The scorer \change{function} is powered by this fine-tuned GPT-4o model. Details on dataset construction and model training are provided in Appendix.

As such, we acknowledge that the scorer is trained on a small, curated dataset. The scoring feedback should be interpreted as an indicative signal for interface interaction and decision-making support, rather than an objective or universal evaluation of the quality of science communication.

To validate the reliability of the scoring mechanism, we conducted a technical evaluation comparing the accuracy of fine-tuned and non-fine-tuned scorers in simulating audience ratings. As shown in \textbf{Table~\ref{fine_tune}}, the fine-tuned scorer exhibited much higher agreement with human ratings (r=0.90/0.91, RMSE$\approx$6--7) than the non-fine-tuned model (r=0.84/0.57, RMSE=22--31). Detailed evaluation is provided in Appendix.

\begin{table}[htbp]
\centering
\caption{Evaluation of the similarity between fine-tuned and original GPT-4o models' scores and human scores.}
\label{tab:similarity_analysis}
\begin{tabular}{lcccc}
\toprule
\multirow{2}{*}{\textbf{Model}} & \multicolumn{2}{c}{\textbf{Pearson Correlation}} & \multicolumn{2}{c}{\textbf{RMSE}} \\
\cmidrule(lr){2-3} \cmidrule(lr){4-5}
 & Engagement & Exposition & Engagement & Exposition \\
\midrule
w/ FT & \textbf{0.90} & \textbf{0.91} & \textbf{6.48} & \textbf{7.02} \\
w/o FT & 0.84 & 0.57 & 22.48 & 30.90 \\
\bottomrule
\label{fine_tune}
\end{tabular}
\end{table}


\textit{Filter \change{function}:} This \change{function} uses the scorer’s outputs to select the top-$k$ versions that best meet the user's expectations. Filter \change{function} ensures that the selected outputs not only fulfill the intended modification chosen direction (Labels) and achieve high scores but also filter out generated failures and low-quality content. This prevents content redundancy and enhances overall generation quality.

\subsubsection{Reinforcement Module}
Since user iterations form a tree of nodes enriched with valuable data (selected labels, prompts, likes/dislikes, and feedback), we developed an analyzer \change{function} to harness both the explicit and implicit signals from these interactions. The analyzer \change{function} captures behavioral data during the iterative process and uses chain-of-thought prompts to interpret user revision behavior.

{\textit{Analyzer \change{function}:}}
The analysis pursues two main goals: (1) identifying common editing patterns, including stylistic preferences, trade-offs between scientific exposition and narrative engagement, and individual user strengths or weaknesses; and (2) uncovering alternative or underused strategy directions. These insights are passed to the Muse component (Section~\ref{Muse}). After the user provides feedback on the LLM’s suggestions through Muse, the Analyzer \change{function} incorporates this real-time feedback (e.g., approvals or further edits) and updates the Recommender \change{function} accordingly. This process refines subsequent strategy recommendations, ensuring that each iteration aligns more closely with the user’s preferences and habits. The feedback loop enables the system to adapt continuously to personal writing habits while balancing narrative engagement and scientific exposition throughout the revision process.

\subsubsection{Implementation}
\system{} is implemented as a web application, with a Python-based backend developed using Flask\footnote{\url{https://flask.palletsprojects.com/en/stable/}} framework and a frontend built using ReactFlow\footnote{\url{https://github.com/wbkd/react-flow/}}.

For the LLM \change{function}s, we employ different LLMs tailored to their functional roles. The recommender, generator, and filter \change{function}s are powered by the GPT-4o-mini model, optimized for fast, high-quality content generation. The analyzer \change{function}, which requires deeper reasoning to interpret user behavior and editing patterns, is supported by the o1 model, which is a reasoning-oriented LLM. The scorer function is powered by a fine-tuned GPT-4o model using a small-sample learning strategy\footnote{\url{https://platform.openai.com/docs/guides/fine-tuning?utm_source=chatgpt.com}}. The frontend communicates with the remote LLMs to obtain results with pre-defined prompt templates. This modular design allows us to tailor \change{function} behavior based on context while maintaining flexibility in prompt construction and LLM selection. The detailed use of prompts in the backend can be found in Appendix.

%% file: sections/User_Study.tex
\change{To better understand how \system{}'s integrated revision environment shapes writers’ cognition and human-AI collaboration during the LLM-assisted science communication writing process, we conducted a controlled user study comparing \system{} with a baseline LLM-supported editing workflow. Our goal was to examine how a revision environment centered on spatial externalization, together with strategy guidance, version organization, and reflective support, influences how writers reason, reflect, and iterate during revision, and to derive design insights for interfaces that better support complex revision processes in co-creation with AI. Accordingly, our evaluation focuses on process-level cognitive and reflective outcomes during revision, rather than direct assessment of whether final texts achieved participants’ intended communication goals.}

\change{\textit{\textbf{RQ1:} How does an integrated revision environment centered on spatial externalization shape users’ cognitive processes during LLM-assisted iterative revision?}}

\change{\textit{\textbf{RQ2:} What interaction tensions and user expectations arise in revision environments with externalized supports?}}

\subsection{Participants}
\change{Rather than representing professionally trained science communicators, our participants reflect a growing group of experienced but non-professional science communication creators with relevant domain knowledge and prior experience producing science-facing content, but without formal training in science communication.} To support this, we recruited 16 participants (9 male, 7 female; aged 24–31, M = 26.9, SD = 2.0), all of whom held postgraduate degrees or higher. Many participants were PhD students, postdoctoral researchers, or early-career faculty affiliated with a local university. The demographic information of these participants is in Appendix.

This study was reviewed and approved by the Institutional Review Board of City University of Hong Kong (IRB No. HU-STA-00001957). All participants provided informed consent prior to participation and were compensated for their time.

\subsection{Procedure}
Each study session began with a live demonstration of the system. Participants were encouraged to explore the interface, try out features, and ask questions. During this walkthrough, the task objectives were also explained.

Each participant completed four text editing tasks: two using the \system{} system and two with the baseline. The texts were selected to represent two common styles of science communication: expository (e.g., “How mRNA Vaccines Work,” “Criteria for Animal Domestication”) and narrative storytelling (e.g., “Discovery of Archimedes’ Principle,” “Living and Thriving with ADHD”). Participants were asked to imagine two specific scenarios: (1) for the expository text: “I have a scientific narrative. How can I make it more engaging and interesting for an online science video?” (2) for the narrative storytelling text: “I have a story as an online science video narrative. How can I link it with more scientific concepts and add scientific credibility?” These two scenarios reflect two common starting points in science communication practice: revising from academically oriented, exposition-heavy scientific content, and developing science narratives from everyday experiences or popular media contexts~\cite{Downs2014PrescriptiveNarratives}.

The length of each text averaged 297.75 words (SD = 19.64). To ensure balanced exposure and mitigate order effects or personal topic preferences, we counterbalanced both the system order (\system{} vs. baseline) and the text type assigned to each system. Thus, each participant edited one expository and one narrative text under each system condition.

Throughout the tasks, participants were encouraged to think aloud, verbalizing their thoughts, reasoning, and feelings as they interacted with the systems. All sessions were screen-recorded, and system interaction logs—such as button clicks (e.g., label selections, generate, regenerate, prompt input, combine)—were automatically captured for the \system{} condition.

\change{The baseline system used in this study consisted of a text editor and a conversational interface (powered by GPT-4o) that supported inline editing and LLM-based suggestions. We selected a common linear editor-and-chat workflow as a conservative comparison because it reflects a realistic and widely used form of LLM-assisted revision practice. In the baseline condition, participants were additionally provided with an Excel table containing strategy names, definitions, usage instructions, examples, and corresponding labels. They were encouraged to use this table as both a conceptual reference and a practical prompting resource by adapting or copy-pasting relevant content into the prompt area during revision. The content in this Excel table was drawn from the same underlying strategy prompts operationalized in \system{}. By contrast, in \system{}, comparable strategic support was embedded directly into the interface through externalized rhetorical goals, actionable labels, and revision-state organization, rather than being accessed as a separate external document.}

\change{As such, the comparison was intended to contrast a familiar linear workflow supplemented by external strategy references with an integrated revision environment that coordinates goals, strategies, and revision trajectories within the interface. Therefore, the study should be interpreted as comparing two different configurations of LLM-assisted revision support.}

\subsection{Post-Task Survey and Instruments}
After completing both conditions, participants completed a post-task survey with standardized instruments: the System Usability Scale (SUS) \cite{brooke1996sus}, NASA-TLX for workload \cite{hart1986nasa}, and the Creativity Support Index (CSI)~\cite{cherry2014quantifying}, with one item adapted to: “I think this system supported me in developing ideas or text collaboratively.” We also asked participants to evaluate the usefulness of the main design features of \system{} using eight questions. 

Besides, we developed a concise co-creation survey targeting two metacognitive constructs from cognitive psychology~\cite{flavell1979metacognition, schraw1994assessing}. Metacognitive knowledge assessed awareness of cognitive goals (e.g., “I am aware of my writing goals during the editing process”). Metacognitive regulation captured planning, monitoring, and evaluation~\cite{qin2022questionnaire} (e.g., “I set specific goals for the narrative,” “I reflect on editing strategies while using the AI tool,” and “I reviewed the narrative to assess how well it communicated scientific content”). These items were adapted from the Metacognitive Awareness Inventory~\cite{schraw1994assessing} and aligned with recent insights into AI-induced metacognitive demands. To measure perceived control during co-creation, we included items inspired by Human-AI interaction principles~\cite{wang2019designing}, focusing on participants’ influence over outputs and narrative direction. Perceived autonomy was assessed according to Self-Determination Theory~\cite{deci2012self}, addressing decision-making freedom, expressive latitude, and resistance to system pressure. The full list of items on metacognition, perception of control and autonomy is provided in Appendix.

All instruments (NASA-TLX, SUS, CSI, and co-creation survey) employed a 7-point Likert scale. After task completion, each participant joined a 15-minute semi-structured interview designed to capture deeper insights into cognitive processes, feature usage, perceived system value, and moments of difficulty or breakthrough. These interviews complemented survey responses and enriched our understanding of user experience across both conditions.

%% file: sections/Results.tex
\change{\subsection{\textbf{RQ1:} How does an integrated revision environment centered on spatial externalization shape users’ cognitive processes during LLM-assisted iterative revision?}}

\change{Drawing on Kirsh’s theory of \textit{thinking with external representations}~\cite{kirsh2010thinking}, we analyze how the integrated revision environment in \system{} supported users’ cognitive processes during LLM-assisted revision. As summarized in \textbf{Table~\ref{tab:thinking_with_external_representations}}, participants described iterative revision as becoming more externally organized and navigable, supporting goal orientation, trajectory-based metacognitive control, and lower-cost exploration.}

\change{These findings should be interpreted as evidence of support for process-level goal awareness and regulation during revision, rather than direct evidence of goal attainment or communication effectiveness.}

\begin{table*}[!t]
\centering
\small
\renewcommand{\arraystretch}{1.25}
\begin{tabular}{
p{2cm}
p{2cm}
p{2cm}
p{4.0cm}
p{4.2cm}
}
\toprule
\textbf{Type of Externalized Support} &
\textbf{Interface Feature} & 
\textbf{Cognitive Function (Kirsh~\cite{kirsh2010thinking})} & 
\textbf{Observed Reasoning and Behavior} & 
\textbf{Representative Evidence} \\
\midrule

\multirow{2}{=}{\textbf{Rhetorical Goals}}
& \textbf{2D coordinate space}  
(scientific exposition $\times$ narrative engagement) 
& \textit{Persistent referents; re-representation of abstract goals}
& Externalized rhetorical trade-offs as a stable design state that users could continuously reference, helping them remain oriented to competing goals and avoid drifting into single-direction revisions 
& “The coordinate graph keeps me from getting lost balancing the two dimensions during revisions” (P3); “I refer to the scores to decide which dimension I need to improve—otherwise I might just keep revising in one direction without noticing as I do in baseline” (P12). \\

& \textbf{Strategy labels aligned with axes}
& \textit{Explicit encoding of strategies; action scaffolding}
& Made rhetorical goals actionable by mapping abstract intentions to concrete revision moves; helped users recognize available strategies and reduced the effort of deciding how to revise 
& “The labels make me realize what kinds of things I should be doing instead of getting lost in details” (P1); “It gave me methods I hadn’t considered before” (P12); “The strategies are packaged—I just click and go” (P7). \\

\midrule

\multirow{2}{=}{\textbf{Iterative Revision Trajectories}}
& \textbf{Node-based version layout with visible scores}
& \textit{Reduced inferential cost; calibration through comparison}
& Enabled side-by-side comparison across multiple versions; scores functioned as indicative reference points to support judgment and prioritization rather than optimization toward a single metric 
& “I can see strengths and weaknesses by comparing the score of different nodes, not just reading one version” (P8); “Now I first check whether the engagement score is higher compared with previous nodes before reading carefully” (P10); “Coordinate scores help me align edits with my standards and visually track progress. Seeing engagement scores rise reinforces my decisions and makes me feel that I am heading in the right direction.” (P3). \\

& \textbf{Persistent revision traces with spatial movement}
& \textit{Trajectory-based reasoning; lowering control cost}
& Supported reflection across iterations by making revision history visible as a trajectory of movement, allowing users to interpret progress, regression, and compensatory adjustments between goals 
& “I can see where each step leads and go back to earlier versions” (P2); “Each version becomes a reference point rather than something I have to remember” (P13); \textit{Behavioral:} Iterative shifts across axes observed in \textbf{Figure~\ref{Iteration_Process}}. \\

\midrule

\textbf{Exploratory Space}
& \textbf{Spatial Parallel Prototyping workspace}
& \textit{Changing cost structure of exploration}
& Lowered the cost of experimentation by enabling non-linear branching, parallel exploration, and reversible decisions without committing to a single path 
& \textit{Survey:} Higher CSI Exploration and Enjoyment scores; “It gave me room to play and test different directions with low cost” (P11); “I can try several versions of editing direction and still come back to earlier ones to make editing in another direction” (P6); “By selecting different labels, I can explore multiple revision directions, while adjusting strategies or prompts to personalize the edits. This gives me a strong sense of creative flexibility.” (P1). \\

\bottomrule
\end{tabular}

\caption{How system interface design supports thinking with external representations~\cite{kirsh2010thinking}, linking spatial externalization features to cognitive functions and observed reasoning behaviors in LLM-assisted revision.}
\label{tab:thinking_with_external_representations}
\end{table*}

\subsubsection{\change{Externalized Rhetorical Goals Help Participants Stay Oriented During Revision}}
Externalizing scientific exposition and narrative engagement as persistent visual dimensions helped participants remain oriented to competing rhetorical goals throughout revision. Rather than reasoning about balance implicitly or retrospectively, participants treated the coordinate space as a stable reference state that made trade-offs continuously visible (\textbf{Table~\ref{tab:thinking_with_external_representations}}, Row 1). This reduced goal drift commonly observed in prompt-only workflows and supported focused prioritization during editing (P1, P8, P12). 

Strategy labels further operationalized these goals by encoding abstract intentions into actionable revision moves, helping users, especially less experienced writers, decide how to revise rather than just whether to revise (P1, P4, P7, P12) (\textbf{Table~\ref{tab:thinking_with_external_representations}}, Row 2). Among all evaluated features, the two-axis feedback (M = 5.94, SD=1.18) and the strategy labels (M = 5.81, SD=1.17) were perceived as the most useful, receiving the highest mean ratings with relatively low variance, highlighting their central role in supporting users’ revision decisions (Appendix \textbf{Figure 12}).

\subsubsection{\change{Externalized Revision Trajectories Supported Metacognitive Control Across Iterations}}
Beyond moment-to-moment orientation, spatial externalization supported metacognitive control across iterations by visualizing the revision trajectory through externalizing the available choices and decisions. Quantitatively, participants using \system{} reported significantly higher levels of metacognition in reflecting on their own strategies and adjusting strategies during editing (Q3, Q4; see \textbf{Figure~\ref{fig:Cognitive}}). 

Participants framed revision as a trajectory-based process, deliberately advancing toward one rhetorical goal and then compensating toward the other to restore balance (as shown in Appendix \textbf{Figure 11}
and \textbf{Figure~\ref{Iteration_Process}}). This behavior reflects metacognitive control, as writers monitored the effects of prior edits and adjusted subsequent strategies accordingly. Rather than treating generations as isolated outputs, they used externalized cues to track revision states over time and coordinate strategy shifts across iterations, \change{enabling reflective revision that remained oriented to externalized rhetorical goals.}

\begin{figure*}[t]
  \centering
  \includegraphics[width=1\linewidth]{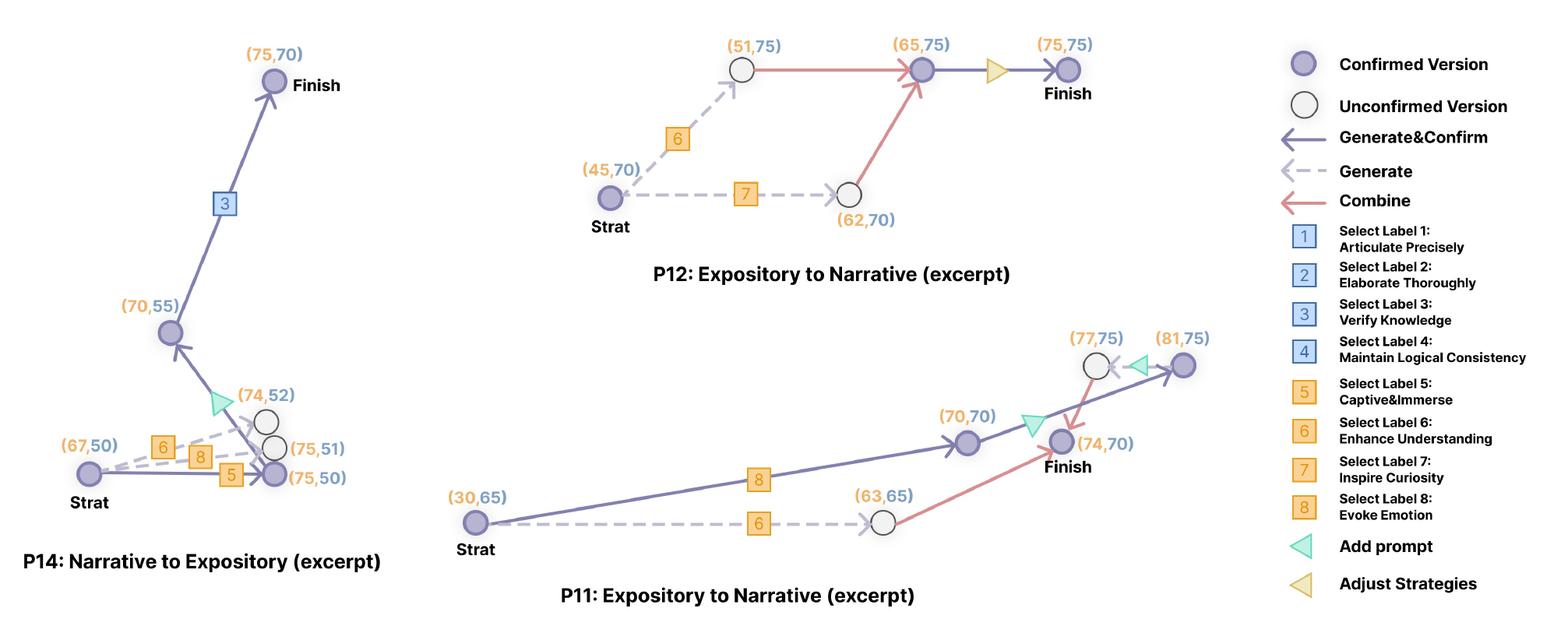} 
  \caption{Visualization examples of segment revisions from P11, P12, and P14.}
  \label{Iteration_Process}
  \Description{Examples of revision trajectories for three participants, labeled P11, P12, and P14. Each trajectory is plotted in a two-dimensional space of narrative engagement and scientific exposition, with points labeled by score coordinates. Purple filled circles indicate confirmed versions and white circles indicate unconfirmed versions. Arrows and line styles distinguish actions such as generating, confirming, combining versions, adding prompts, and adjusting strategies. The three examples show different revision paths: one moves from narrative-heavy toward stronger exposition, while the others move from expository starting points toward more narrative versions. Together, the figure illustrates how participants revised by branching, combining, and iteratively shifting between the two rhetorical goals rather than following a single linear path.}
\end{figure*}

Qualitatively, the node-based version layout with visible scores and the persistent revision traces with spatial movement jointly supported decision making and process-level control during iterative revision (\textbf{Table~\ref{tab:thinking_with_external_representations}}, Rows~3–4). By enabling side-by-side comparison across versions, visible scores reduced inferential cost and provided indicative reference points that helped participants judge relative strengths, prioritize revision directions, and decide where to invest attention (P1, P8, P10, P3). Beyond local decisions, persistent spatial traces externalized revision history as a trajectory, allowing participants to interpret progress, regression, and compensatory shifts between rhetorical goals (P2, P13).

Furthermore, scores were used for calibration rather than optimization, reinforcing confidence in the revision process. Just as P3 mentioned, “Coordinate scores help me align edits with my standards and visually track progress. Seeing
engagement scores rise reinforces my decisions and makes me feel that I am heading in the right direction.” By making progress perceptible across iterations, externalization reduces epistemic uncertainty about whether local edits contribute to higher-level goals, thus making participants feel more confident. 

\subsubsection{\change{The Externalized Workspace Lowered the Perceived Cost of Exploration}}

Externalizing the exploratory space also supports creativity. Participants rated \system{} significantly higher in Exploration and Enjoyment on the CSI questionnaire (\textbf{Figure~\ref{fig:CSI}}), without increases in perceived cognitive load (NASA-TLX; \textbf{Table~\ref{fig:NASA_and_SUS}}). Qualitative insights indicate that the shared spatial workspace enabled non-linear branching, parallel comparison, and reversible decisions, lowering the risk and effort associated with experimentation (\textbf{Table~\ref{tab:thinking_with_external_representations}}, Row 5). The free exploratory space also allowed users to explore multiple revision directions simultaneously, encouraging playful testing and occasional conceptual shifts that would be less likely in linear prompt-response workflows.

\begin{table*}[h]
\centering
\begin{tabular}{l|lcccccc}
\hline
& & \multicolumn{2}{c}{\system{}} & \multicolumn{2}{c}{Baseline} & \multicolumn{2}{c}{Statistics} \\
\cline{3-8}
& & mean & std & mean & std & p-value & Sig. \\
\hline
\multirow{6}{*}{\rotatebox[origin=c]{90}{\textbf{NASA-TLX \cite{hart1986nasa}}}}
& Mental Demand & 4.63 & 1.36 & 4.19 & 1.68 & .404 & — \\
& Physical Demand & 3.19 & 1.60 & 2.63 & 0.96 & .261 & — \\
& Temporal Demand & 2.63 & 1.36 & 3.19 & 1.38 & .343 & — \\
& Effort & 3.94 & 1.39 & 4.44 & 1.79 & .241 & — \\
& Performance & 5.13 & 0.89 & 4.88 & 0.96 & .372 & — \\
& Frustration & 2.88 & 1.59 & 3.00 & 1.32 & .724 & — \\
\hline
\multirow{10}{*}{\rotatebox[origin=c]{90}{\textbf{SUS \cite{brooke1996sus}}}}
& Q1: use frequently & 5.13 & 1.54 & 4.38 & 1.36 & .155 & — \\
& Q2: unnecessarily complex & 3.00 & 1.41 & 2.94 & 0.85 & .899 & — \\
& Q3: easy to use & 4.94 & 1.69 & 4.88 & 1.15 & .964 & — \\
& Q4: need support & 3.94 & 1.91 & 2.81 & 1.87 & .031 & $*$ \\
& Q5: function well integrated & 5.13 & 1.26 & 3.44 & 1.36 & .003 & $**$ \\
& Q6: inconsistency & 3.06 & 1.39 & 3.25 & 1.53 & .719 & — \\
& Q7: learn to use quickly & 4.88 & 1.59 & 5.06 & 1.44 & .604 & — \\
& Q8: awkward & 2.44 & 1.26 & 2.50 & 1.37 & .927 & — \\
& Q9: confident & 4.50 & 1.32 & 4.50 & 1.37 & .812 & — \\
& Q10: need learning & 3.81 & 1.56 & 3.38 & 1.89 & .397 & — \\
& Overall Score & 70.78 & 29.70 & 68.44 & 26.94 & .729 & — \\
\hline
\end{tabular}
\caption{The statistical results of NASA-TLX and SUS questionnaires. ($*$: $p<0.05$ and $**$: $p<0.01$).}
\label{fig:NASA_and_SUS}
\end{table*}

\change{\subsection{RQ2: What interaction tensions and user expectations arise in revision environments with externalized supports?}} 

\subsubsection{Balancing Externalized Guidance and User Judgment}
Participants described how the system's visual and scoring feedback may influence their evaluation practices in subtle ways. While the coordinate axis enabled intuitive comparisons between revisions, some participants noted that the visibility and immediacy of scores could reduce their depth of textual engagement. As P4 reflected, “When using the system, I outsourced a large part of the thinking process to the AI. It's faster and more efficient, but I also tend to think less carefully about the output as I trust the score results more than I did with the baseline. In baseline, I would read text more carefully and make judgments by myself.” This suggests that while externalized scoring streamlines comparison, it can also shift evaluative effort away from close reading toward greater reliance on system-provided judgments.

Others expressed a degree of caution about over-relying on the scores. P16 noted that while the visual feedback was useful, “The scores are indicative rather than definitive. They sometimes do not reflect the actual quality of the generation and still require human judgment.” P7 also noted that although the coordinate view provides scores, they still read the text carefully and reconcile the system’s feedback with their own standards. As a result, they sometimes chose versions located at intermediate positions rather than pursuing extreme scores. These reflections suggest a potential tension: while the system offers accessible and actionable feedback, its effectiveness depends on users' ability to critically interpret the signals rather than accept them at face value. 

Concerns about the interpretability of scoring were also raised. As P14 said, “Sometimes I don't know what an increase in score actually means. I can't tell whether each label contributes differently to the score or what specific content led to a higher score. I want to understand the logic behind the numbers.” This suggests the interpretability of the scores and the changes made to them also needs improvement. 
\begin{figure*}[t]
  \centering
  \includegraphics[width=\linewidth]{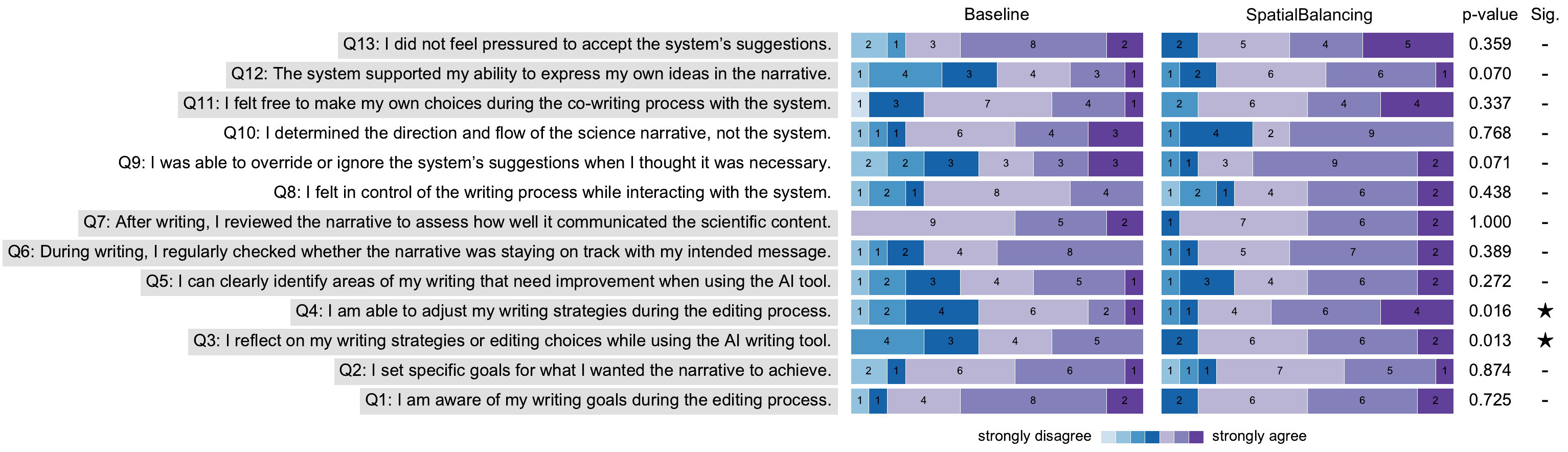}
  \caption{Results of the Metacognition (Q1–Q7), Control (Q8–Q10), and Autonomy (Q11–Q13) questionnaires (p < .05 marked with *; p < .01 with **). Significant differences were observed in Metacognition: Q3 (M = 5.50 (\system{}) vs. 4.63 (Baseline), p = .013) and Q4 (M = 5.69 vs. 4.56, p = .016); marginal differences in Control: Q9 (M = 5.63 vs. 4.75, p = .071) and Autonomy: Q12 (M = 5.25 vs. 4.44, p = .070).}
  \label{fig:Cognitive}
  \Description{Comparison of questionnaire responses between the baseline condition and Spatial Balancing across 13 items grouped into metacognition, control, and autonomy. Each row is a stacked bar chart on a seven-point agreement scale, ranging from strongly disagree to strongly agree. The left column shows baseline responses and the right column shows Spatial Balancing responses, with p-values and significance markers listed at the far right. Most items show similar distributions across conditions, but Spatial Balancing has significantly higher ratings on two metacognition items: reflecting on writing strategies while using the AI tool and adjusting writing strategies during editing. The caption also notes marginal differences for feeling able to express narrative ideas and making one’s own choices during co-writing.}
\end{figure*}

\begin{figure}[t]
  \centering
  \includegraphics[width=\linewidth]{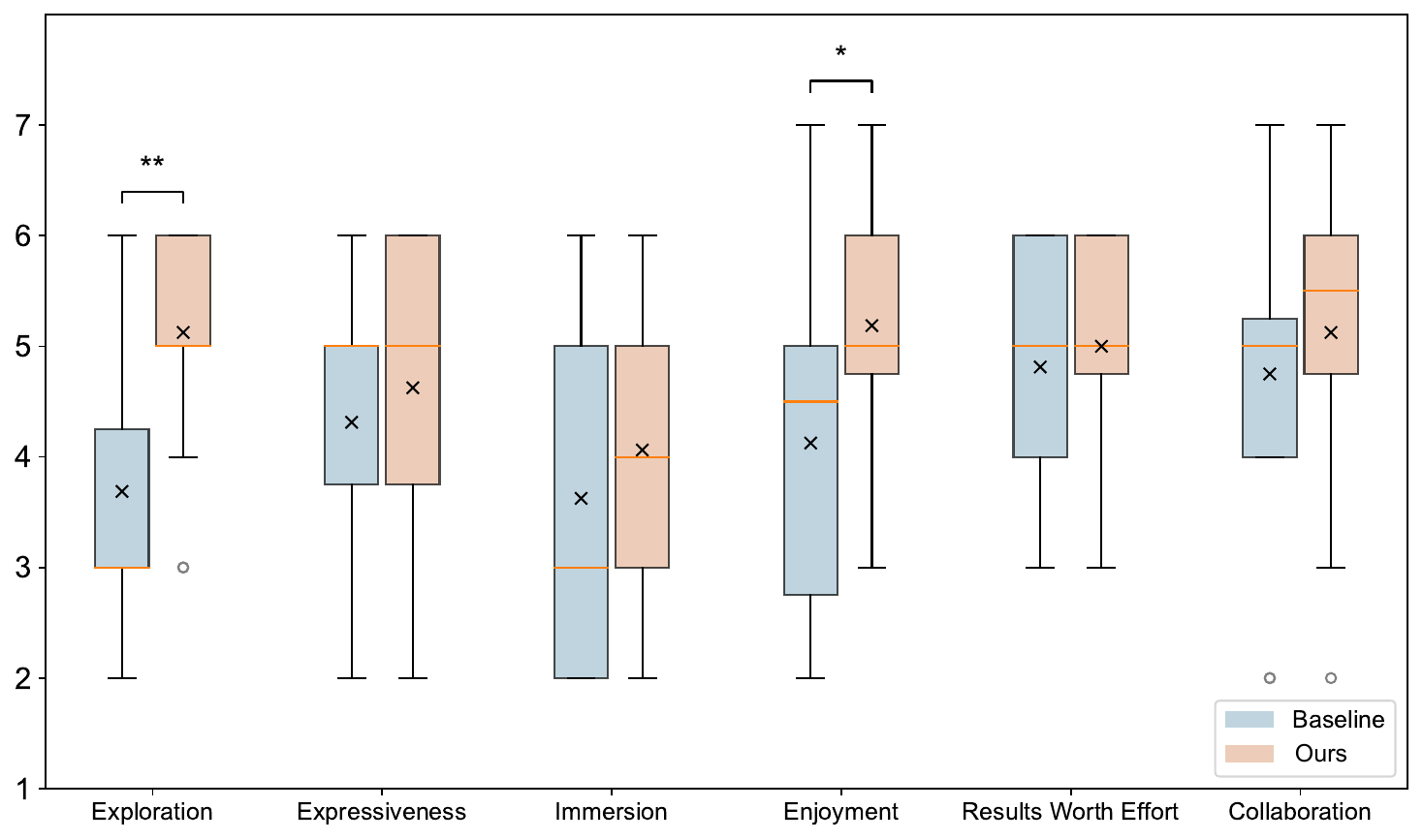}
  \caption{The results of CSI questionnaire. ($*$: $p<0.05$ and $**$: $p<0.01$). Participants rated \system{} significantly higher in terms of “Exploration” (M = 5.13 (\system{}) vs. 3.69 (Baseline), p = .004) and “Enjoyment” (M = 5.19 vs. 4.13, p = .039)}
  \label{fig:CSI}
  \Description{Box plots comparing Creativity Support Index responses for the baseline condition and Spatial Balancing across six dimensions: exploration, expressiveness, immersion, enjoyment, results worth effort, and collaboration. Each pair of box plots shows the distribution of participant ratings on a seven-point scale, with means marked by x symbols and significance indicated above some categories. Spatial Balancing is rated significantly higher than the baseline on exploration and enjoyment, while the other four dimensions show similar distributions between conditions.}
\end{figure}

\subsubsection{Seeking More Flexibility and Adaptivity of Externalization in Use}
While the eight-label set was seen as a helpful starting point, more experienced participants felt it could be more flexible and customizable to better support their advanced needs. Several participants (P1, P3, P2, and P14) wished they could combine or tailor underlying strategies to form customized labels to align more closely with their specific intentions. P1 expressed a desire to curate combinations of strategies based on their own habits, and to flexibly create new combinations to support more personalized needs. P14 also noted, “In addition to the current style-focused labels, it would be helpful to include others that target areas in writing revision like grammar or tone.” Together, these responses reflect a tension between predefined externalized guidance and users’ desire for greater agency.  
 
Participants also reported that repeated exposure to the coordinate scores helped them develop a personal reference range, allowing them to recognize patterns in their own writing habits over time (P3, P4). Rather than treating the scores as absolute targets, they used them to understand where their typical writing tended to fall and how revisions shifted that position. As P3 suggests, “If the visualization can provide further visual indication of the score ranges preferred by specific reader groups, it could enable more informed adjustments by helping authors intentionally move the revision points toward positions that better align with different audience expectations.” This indicates that participants appropriated the coordinate scores as a personalized, evolving reference system rather than fixed evaluative benchmarks, using repeated exposure to calibrate their own writing tendencies and reason about audience-specific adjustments.

\subsubsection{More Proactive and Grounded Feedback in the Revision Process}
While participants appreciated what Muse could already do to help reflect on the whole revision process (P2, P6, P13, P14), P2 wanted more real-time dialogue: “I wish it were more interactive—like chatting with someone who helps me reflect as I go during the revision process.” P13 and P14 also expected the system to proactively offer assistance, even before they explicitly recognized the need for help. The log data further indicates that participants tended to use the Muse function primarily at the final stage of their revision and only once in most cases (Appendix \textbf{Figure 11}). 
This points to the need for more proactive and embedded reflective interactions rather than relying on users to initiate reflection themselves.

Participants also wished that Muse could give more personalized and context-specific feedback in the revision process (P1, P8). “Right now, Muse gives high-level suggestions,” P8 said. “But it'd be more useful if it could point to which step or decision was strong or weak, and explain why.” This suggests that participants seek feedback that is grounded in specific revision actions and their underlying rationale.

%% file: sections/Discussion.tex
\subsection{Design Insights for Externalization in Human-AI Writing Interfaces}
\subsubsection{Design Insight 1: Mitigating Metacognitive Laziness of Relying on Externalization}
According to distributed cognition, reflection and problem solving are not confined to an individual’s internal reasoning, but are distributed across interactions among users, AI models, and external tools~\cite{10.1145/353485.353487}. In such distributed cognitive systems, merely providing access to knowledge or suggestions is insufficient. Users must also be able to understand, monitor, and regulate how this knowledge is produced, interpreted, and applied within the system—an ability that lies at the core of metacognition~\cite{Sidra19082025}. Our user study suggests that externalizing rhetorical goals and revision trajectories can effectively enhance metacognitive regulation during LLM-assisted revision. By making abstract goals and revision progress perceptible through spatial cues, users were better able to reflect on their editing strategies, adjust revision directions across iterations, and maintain a sense of process-level control. These findings align with prior work showing that external representations can support planning, monitoring, and evaluation by reducing the inferential burden of tracking change internally~\cite{kirsh2010thinking}.

However, the same externalization properties also reveal a potential tension. Highly legible and actionable feedback—such as explicit scores and spatial comparisons—can simultaneously support reflection and displace reflective effort. Several participants reported that they began to rely more on system-provided cues to make decisions, sometimes at the expense of close reading and independent evaluation of the text. In these moments, evaluative judgment shifted from users’ own critical reasoning toward system-generated signals. This echoes prior findings that frequent reliance on LLM feedback may encourage over-trust and lead to “metacognitive laziness,” in which users reduce self-regulation and critical engagement with the task~\cite{https://doi.org/10.1111/bjet.13544,Sidra19082025}.

Together, these findings highlight an important design challenge: rather than treating cognitive offloading as an unqualified benefit, designers should carefully consider what aspects of cognition are externalized and how users are invited to engage with them~\cite{Sidra19082025}. Design considerations for using these kinds of externalized visualization features include: (1) Designing feedback as reflective prompts rather than prescriptive, for example, by framing scores as indicative signals that invite interpretation, comparison, or questioning, instead of optimization targets; (2) Supporting moments of deliberate re-engagement, such as encouraging users to articulate why they accept, reject, or override system suggestions, thereby reinforcing evaluative ownership; (3) Providing adjustable levels of guidance, allowing users to control when and how much evaluative feedback is visible, so that reliance on external cues can be modulated over time and expertise levels. (4) Making the basis of system feedback more interpretable, helping users understand why certain revisions shift scores, which can transform externalized metrics from authority signals into learning resources.

\subsubsection{Design Insight 2: Preserving Agency through Adaptive Mixed-Initiative Externalization}
Scaffolding through externalization is effective for supporting rapid prototyping and reducing decision overhead in LLM-assisted writing, particularly in early stages of revision. By packaging strategies into higher-level labels, the system helps users quickly explore and compare revision directions. However, as users gain experience, fixed scaffolds can become constraining, no longer aligning with their evolving intentions, personal writing habits, or situational goals. Our findings show that experienced users wanted to move beyond predefined labels by curating and recombining underlying strategies, treating externalized structures not as fixed guidance but as resources to be reshaped.

These findings suggest that externalization should function as a flexible and adaptive substrate, rather than a static scaffold. Interfaces should support user-driven customization of externalized elements (e.g., allowing users to create or curate personalized labels), while also enabling system-driven adaptation based on observed interaction patterns. For example, by reflecting stable writing patterns back into the visualization—such as indicating personal reference zones or audience-specific target regions within the exploratory space—the system can adapt externalized cues to users’ evolving goals and habits. In this way, externalization shifts from prescribing ideal targets to supporting situated self-regulation. Previous work suggests that agency in human-AI co-creation
fluctuates across the creative process~\cite{rafner2025agency,han_when_2024}, so designs that adapt externalization in this manner preserve the efficiency benefits of scaffolding while gradually restoring user agency, enabling more individualized, reflective, and sustainable writing practices in human-AI collaboration.

\change{Our findings also suggest that flexibility should apply not only to how guidance is delivered, but also to how rhetorical goals are represented. While the predefined two-dimensional space served as a useful scaffold in our study, future systems should explore more negotiable goal representations that better reflect writers’ situated and evolving priorities—for example, allowing writers to temporarily swap one rhetorical dimension for a task-specific concern such as tone, audience appropriateness, or platform fit.}

\subsubsection{Design Insight 3: Providing Proactive In-Situ Reflective Support}
While reflective support is essential for helping writers make sense of iterative revisions, relying on users to explicitly initiate reflection can limit its effectiveness. In our results, we found that when reflection relied primarily on user initiative, participants were less likely to engage in it proactively. Instead, they expressed a preference for more step-by-step, in-situ reflective support integrated into the revision process. 

This finding points to the need for proactive mechanisms that surface reflective support at appropriate moments within the revision process. For example, rather than relying on users to pause and reflect on their own, systems should proactively trigger reflection at natural breakpoints in revision, such as when users compare alternatives, confirm a revision, or shift revision direction, instead of expecting users to stop. In addition, reflective support should be grounded in visible artifacts of revision~\cite{10.1145/3746059.3747703,10.1145/3706598.3714316} such as generated alternatives, score changes, or spatial movements—to make reflection concrete and interpretable. For example, when users repeatedly explore multiple versions of a passage without reaching a satisfactory outcome, the system can proactively surface reflective questions besides the revision node to help clarify underlying intentions.

\subsection{Limitations and Future Work}

We describe several limitations in the study to define the scope of our findings clearly and motivate future work.

\subsubsection{\change{Limits of the Current Baseline and Directions for Future Comparison}}

\change{A limitation of the current study is that the baseline captured a realistic editor-and-chat workflow, but did not fully align with \system{} in how revision support was structured, invoked, and presented during use. In \system{}, guidance, alternatives, scores, and revision history were integrated into the interface, whereas in the baseline these forms of support were more externally referenced and manually assembled. As a result, the current comparison is better understood as contrasting two different configurations of revision support, rather than as strictly isolating the effect of spatial externalization alone. Accordingly, the present findings should be interpreted as evidence about the benefits and tensions of an integrated revision environment, rather than as a definitive causal test of spatial externalization in isolation. Future studies could isolate the contribution of individual features, such as goal-space visualization, score feedback, and revision traces, through targeted ablation or feature-level comparisons to better identify which aspects of the environment most strongly shaped users' revision processes.}

\change{Although the baseline Excel materials were derived from the same underlying strategy content used in \system{}, this did not ensure equivalence in how LLM support was enacted during revision. In \system{}, these supports were embedded into the interaction flow, whereas in the baseline, participants had to manually consult and re-enter them through chat. Future studies should therefore adopt a more harness-equivalent comparison that preserves the same underlying supports while varying only their interface presentation.}

\subsubsection{\change{Lack of Outcome-Based Evaluation of Goal Attainment, Text Quality, and Communication Effectiveness}}
\change{A key limitation of the current study is that, although we examined how the system supported writers’ goal awareness, strategy reflection, and metacognitive regulation during revision, goal achievement was assessed primarily through participants’ subjective perceptions during the process. We did not independently evaluate whether the final texts ultimately achieved their intended communication goals. Accordingly, our findings should be interpreted as evidence of support for goal-aware iterative revision, rather than outcome-based evidence of goal attainment. In addition, we did not systematically assess the quality of the resulting texts or their communication effectiveness, such as how they were received by intended audiences or whether they improved reader understanding or engagement. Future work should combine participants’ self-assessments with expert review and audience-based evaluations.}

\subsubsection{Evaluation Dependency on Proxy Scores}
To demonstrate \system{} with minimal evaluation overhead, we adopted a low-cost approach that uses model-generated proxy scores to approximate audience feedback on scientific exposition and narrative engagement. These proxies were intended to support comparative reasoning and iterative decision making during revision, rather than to represent comprehensive or definitive audience judgments. While such scores may reflect the expectations of a particular participant group, they cannot capture the full diversity of real-world audiences or contexts (e.g., classroom learning vs.\ online videos). Accordingly, the current scoring mechanism should be understood as a design probe, and future work should validate and extend it with audience- and context-specific evaluation methods.

\subsubsection{\change{Tensions of Externalized Scoring as Guidance}}
\change{A further limitation lies not only in what the scores measure, but also in how they guide behavior once externalized in the workspace. In \system{}, the 2D coordinates were intended as navigational cues for reflection and comparison. However, once these cues become persistent and visually prominent, writers may begin to treat them as goals in themselves. This creates a tension: the same externalized signals that help users stay oriented and compare alternatives may also over-shape revision behavior, encouraging score-seeking, over-trust in system judgments, or reduced attention to situated rhetorical concerns. Future work should therefore make such scoring mechanisms more interpretable and negotiable, for example by exposing the basis of placements, supporting user correction, and allowing writers to define more personalized goal regions or rhetorical dimensions.}

\subsubsection{Methodological Limitations.} This work has common methodological limitations including the short-term nature of system testing which may not reveal long-term adoption patterns, and the relatively homogeneous participant demographics that may not represent all potential user groups. Future work will aim to address the previously mentioned and these limitations through more comprehensive evaluations. 

%% file: sections/Conclusion.tex

Our results show that spatial externalization reshapes how writers reason about LLM-assisted revision. By externalizing rhetorical goals and revision history in an exploratory spatial workspace, participants treated revision as a trajectory rather than a series of isolated edits, enabling sustained goal orientation, metacognitive control across iterations, and low-cost exploration of alternatives. The two-dimensional feedback functioned as navigational cues—supporting calibration and reflection, rather than prescriptive optimization signals. At the same time, participants surfaced tensions around over-reliance on externalized scores and the need for more flexible, adaptive forms of externalization. Together, these findings suggest that the value of spatial exploratory interfaces lies not in generating better revisions per se, but in supporting writers’ ability to navigate, reflect on, and steer complex revision processes over time.

%% file: sections/Appendix.tex
\subsection{Specific Strategies for Science Communication Writing}

\begin{table*}[htbp]
\centering
\caption{Design Space for Science Communication Writing}
\label{DesignSpace}
\renewcommand{\arraystretch}{1.1}
\scalebox{1}{ 
\footnotesize
\begin{tabular}{c|p{3cm}|p{7cm}|c}
\hline
\textbf{Category} & \textbf{Strategy} & \textbf{Definition} & \textbf{Label} \\ \hline

 \multirow{10}{*}{\makecell{Scientific \\ Exposition}}&(1) Layered Transitions~\cite{Rossmann2025, konig2024communicate, Maskill1988, Volpato2015}
& Use multiple transition words or phrases (e.g., “but,” “and,” “therefore”) within a short span to emphasize logical shifts and contrasts.& 4 \\ \cline{2-4}

 &(2) Rigorous Source \newline Verification~\cite{augenstein2021determining, roland2009quality, konig2024communicate}
 &Cross-check scientific claims and data against reliable, peer-reviewed sources to ensure exposition.& 3 \\ \cline{2-4} 

 &(3) Step-by-Step \newline Explanation~\cite{konig2024communicate, august2020writing}
 &Introduce the core idea first and then progressively add background details, creating a structured learning process.& 2, 4\\ \cline{2-4} 

 &(4) Acknowledge \newline Uncertainties~\cite{pielke2007honest}
 &Transparently discuss uncertainties, potential biases, or limitations in data and models to build credibility.&1, 2\\ \cline{2-4} 

  &(5) Consistent \newline Terminology~\cite{Kueffer2014}
 &Use the same terminology throughout the content to maintain clarity and avoid confusion.& 1 \\ \cline{2-4} 

  &(6) Citations \& Quotes~\cite{augenstein2021determining, ellis2022improving}
 &Integrate citations and direct quotes seamlessly to enhance credibility while maintaining narrative flow.& 3 \\ \cline{2-4}

 &(7) Everyday Events to \newline Scientific Insights~\cite{august2020writing, Kueffer2014}
 &Automatically identify and link theories or knowledge to real-world events or stories mentioned in the text. &2, 3\\ \cline{2-4} \hline 
\multirow{18}{*}{\makecell{ Narrative\\ Engagement}}& (8) Question-Answer Hook~\cite{finkler2019power,huang2020good,lambert2013digital}&Ask a direct question and provide an immediate answer to introduce key concepts clearly and concisely. & 5, 6, 7 \\ \cline{2-4}

& (9) Reflection Question~\cite{finkler2019power}
 & Ask a thought-provoking question that does not require an immediate answer, encouraging reflection and reinforcing key concepts.& 5, 7, 8 \\ \cline{2-4} 
 
& (10) Suspense-Driven \newline Reveal~\cite{zhang2023understanding,xia2022millions}
 & Present a question, problem, or scenario at the beginning and delay its resolution to sustain curiosity.&5, 7\\ \cline{2-4} 
 
&(11) Use Metaphors~\cite{finkler2019power, Kueffer2014}
 &Convey unfamiliar concepts by drawing analogies to more familiar ones.& 5, 6\\ \cline{2-4} 

&(12) Inject Humor~\cite{Goodwin2014}
 &Use playful language or puns to make the content more engaging and enjoyable.&5, 8\\ \cline{2-4} 

 & (13) Add Real-World Supporting Examples~\cite{lehrman2019political,livo1986storytelling}
 & Illustrate abstract concepts using relatable, real-world examples.& 5, 6\\ \cline{2-4} 

 & (14) Add Stories~\cite{dahlstrom2014using,livo1986storytelling, Dahlstrom2018paradox}
 &Use narratives with characters, settings, and plot progression to enhance engagement and memorability.&5, 6, 8\\ \cline{2-4} 

 &(15) Add an Imagery \newline Description~\cite{NarrativebasedLearningPossible,finkler2019power,andrews2018advertising}
 & Use vivid, sensory details to help the audience visualize concepts.&5, 6\\ \cline{2-4} 

  &(16) Create Negative Emphasis for Focused Attention~\cite{finkler2019power,NarrativebasedLearningPossible,huang2020good,munozmorcilloTypologiesPopularScience2016}
 & Highlight extreme negative outcomes to intensify focus and reinforce key lessons.&  5, 8\\ \cline{2-4} 

  &(17) Make Positive Emotion to Expand Action Repertoire~\cite{NarrativebasedLearningPossible, finkler2019power,garland2010upward,weinreich2010hands,rowe2007positive,munozmorcilloTypologiesPopularScience2016}
 & Use uplifting messages, particularly in conclusions, to inspire optimism and motivation.& 5, 8\\ \cline{2-4}\hline

 &(18) Simplify and Abstract \newline Language~\cite{Kerwer2021, Suto2023, Ivchenko2022}& Rephrase complex scientific terminology or detailed descriptions into more general, accessible language without compromising core exposition.& 1, 6\\ \cline{2-4} 

   &(19) Clarify Key Terms~\cite{munozmorcilloTypologiesPopularScience2016, Rossmann2025}
 &Define complex or specialized terms at the beginning to establish a shared understanding.&1, 6\\ \cline{2-4}
 
&(20) Key Point Recap~\cite{finkler2019power,torresScientificStorytellingNarrative2019, munozmorcilloTypologiesPopularScience2016}
 &Summarize the main points concisely at the conclusion of the content to reinforce memory retention.&1, 4, 6\\ \cline{2-4} 

 \multirow{6}{*}{Both} &(21) Repeat Key Point(s) or Question(s)~\cite{Kaur_2012, Boubertakh_2015}
 &Reinforce key concepts by strategically repeating crucial terms or questions.&1, 6\\ \cline{2-4}

&(22) Emphasize with \newline Numbers~\cite{yang2021design, Fogg-Rogers_Grand_Sardo_2015}
 &Use specific numbers, statistics, or quantitative comparisons to make key points concrete, precise, and memorable.& 1, 2, 3, 8\\ \cline{2-4}

  &(23) Strengthen the Connections Between Content~\cite{Maskill1988, Volpato2015}
 &Ensure smooth transitions between related ideas by using bridging statements or contextual links.& 4, 6 \\ \cline{2-4} 

 &(24) Present Balanced Views~\cite{Kueffer2014}
 &Provide both supporting evidence and counterarguments to present a well-rounded discussion.&2, 6\\ \cline{2-4} 
 
  &(25) Tie Science to Current Events~\cite{august2020writing, Kueffer2014}
 &Connect scientific discussions to real-world recent news or relevant stories.&3, 5, 6\\ \cline{2-4} \hline
\end{tabular}}

\textbf{*Label: }\textit{Scientific Exposition Effects:} 1. Articulate Precisely; 2. Elaborate Thoroughly; 3. Verify Knowledge; 4. Maintain Logical Consistency\newline \textit{Narrative Engagement Effects:} 5. Captivate \& Immerse; 6. Enhance Understanding; 7. Inspire Curiosity; 8. Evoke Emotion 
\end{table*}

\newpage
\subsection{Rating Model Construction}\label{subsec:rating-model}
 
Our primary goal in constructing the coordinate axis is to simulate audience feedback so that users can receive real-time evaluations. Therefore, we collected real user feedback on texts with varying characteristics to fine-tune an LLM that can provide scores during the real-time writing process.

\textit{Dataset Construction} We first built a dataset of popular science texts containing 45 texts from five commonly seen science communication topics: psychology, economics, geography, history, and physics. 
For each topic, there are nine texts; three each of long (300 words), medium (150 words), and short (50 words) formats; representing three typical levels of revision granularity in science communication. Within each length category, we included three different levels of narrative transformation: (1) purely expository scientific texts (Expository), (2) fully narrative story-like texts (Story), and (3) an intermediate “infotainment” style (Medium), which is an ideal format in popular science that maintains scientific exposition while incorporating narrative strategies from our design space. All texts were revised by an expert with two years of experience in science communication writing

\textit{Score Collection} We designed a survey to collect ratings for these texts on two dimensions: Narrative Engagement and Scientific Exposition, two main communication goals in popular science~\cite{dahlstrom2014using}. For Narrative Engagement, we used five subscales: Narrative Presence, Emotional Engagement, Narrative Understanding, Curiosity, and General Narrative Engagement, a survey developed by prior work~\cite{busselle2009measuring}. For Scientific Exposition, given the lack of mature scales, we measured five dimensions inspired by standards for scientific texts from previous research~\cite{dahlstrom2014using}: Conceptual Clarity, Plausibility, Completeness, Perceived Factual Correctness, and General Scientific Accuracy. The full questionnaire can be found in Section~\ref{surveyitems}. 

\textit{Participants}
First, we recruited three experts (each with more than one year of experience in creating science narratives) to rate the texts. After rating, they discussed and jointly established a scoring rubric, including benchmarks for each score range from 0 to 10. Next, we recruited 27 participants interested in science communication. We invite experts to establish standards as a reference point for audience ratings, in order to reduce variance in their subjective evaluations of the text. 

\textit{Survey Results}
The distribution of scores for the 45 texts is displayed in \textbf{Figure~\ref{fig:45}}. It is shown that story-like texts tend to elicit higher narrative engagement but exhibit lower scientific exposition. In contrast, expository texts maintain higher scientific exposition at the expense of engagement. The infotainment style appears to strike a balance between the two. Additionally, longer texts generally perform better in both dimensions, whereas shorter texts show lower overall scores, likely due to limitations in content depth and development.

\vspace{\baselineskip}
\begin{figure}[t]
  \centering
  \input{figs/survey.tex} 
  \caption{Each point represents one of 45 science communication texts, plotted by its average audience rating for narrative engagement (x-axis) and scientific exposition (y-axis), based on 27 crowd-sourced rubric-based evaluations per text. The left panel groups texts by narrative style: Expository (informational, fact-focused), Story (highly narrative), and infotainment (represents infotainment-style revisions that blend factual exposition with narrative strategies). The right panel groups texts by length (Short=50 words, Medium=150 words, Long=300 words).} 
  \label{fig:45}
  \Description{Two scatter plots showing the distribution of 45 science communication texts by average audience ratings on two dimensions: narrative engagement on the horizontal axis and scientific exposition on the vertical axis. In the left plot, points are grouped by narrative style: expository texts cluster toward higher scientific exposition and lower narrative engagement, story texts tend to have higher narrative engagement and more varied exposition, and infotainment texts fall between these patterns by combining elements of both. In the right plot, points are grouped by text length: short texts cluster toward lower values on both dimensions, medium-length texts occupy the middle range, and long texts more often appear at higher scientific exposition and moderate to high narrative engagement. Together, the figure shows how the dataset varies across style and length in the two-dimensional rhetorical space.}
\end{figure}

\textit{Final Model Fine-Tuning}
For each text, we first computed the average score across the five questions within each of the two dimensions and then averaged these scores across all 27 participants. To match the 0–100 scale of the final coordinate axis, the scores were scaled by a factor of 10. These scaled scores (representing the two dimensions) served as the output, while the corresponding text and the expert-defined criteria used as reference formed the input. 

During the development phase, we adopted a small-sample fine-tuning strategy to customize GPT-4o for our domain-specific application. This approach, which leverages a relatively limited number of high-quality training examples, has been shown to be both efficient and practically effective in enhancing model performance on specialized tasks~\footnote{\url{https://platform.openai.com/docs/guides/fine-tuning?utm_source=chatgpt.com}}. We prepared and uploaded the curated dataset through OpenAI’s official platform and used their fine-tuning API to tailor GPT-4o. The resulting customized model served as the backbone of our scoring system.

\textit{Technical Evaluation}\label{Technical_evaluation}
To validate the reliability of this scoring mechanism, we conducted a formal evaluation. We constructed a controlled dataset consisting of five source articles, each systematically rewritten into three different lengths (long, medium, short) and expressed in three different styles (expository, medium, story). This design yields nine distinct variants per article, resulting in a total of 45 text samples.
From this dataset, we randomly selected 33 samples for fine-tuning GPT-4o, while reserving 12 samples for evaluation. The fine-tuned model was assessed against human ratings on two key dimensions: narrative engagement and scientific exposition.
On the held-out test set, the fine-tuned model demonstrated a high degree of alignment with human judgment, achieving Pearson correlation coefficients of 0.90 and 0.91 for narrative and exposition scores, respectively. In addition, the model’s predictive reliability was reflected in RMSE values of 6.48 and 7.02. These results indicate that the fine-tuned LLM scoring mechanism can effectively approximate human evaluative patterns, thereby providing a reliable and scalable alternative to manual scoring.

\subsection{Survey}\label{surveyitems}

\textbf{Part 1: Metacognition}

Metacognitive Knowledge: This pertains to an individual's awareness and understanding of their own cognitive processes and strategies

Q1: I am aware of my writing goals during the editing process.

Strongly Disagree\quad 1\hspace{1em}2\hspace{1em}3\hspace{1em}4\hspace{1em}5\hspace{1em}6\hspace{1em}7\quad Strongly Agree
\vspace{1em}

Metacognitive Regulation: This involves the active management of one's cognitive processes through planning, monitoring, and evaluating

Q2: I set specific goals for what I wanted the narrative to achieve.

Strongly Disagree\quad 1\hspace{1em}2\hspace{1em}3\hspace{1em}4\hspace{1em}5\hspace{1em}6\hspace{1em}7\quad Strongly Agree
\vspace{1em}

Q3: I reflect on my writing strategies or editing choices while using the AI writing tool.
(Indicates real-time assessment of strategy effectiveness.)

Strongly Disagree\quad 1\hspace{1em}2\hspace{1em}3\hspace{1em}4\hspace{1em}5\hspace{1em}6\hspace{1em}7\quad Strongly Agree
\vspace{1em}

Q4: During writing, I regularly checked whether the narrative was staying on track with my intended message.

Strongly Disagree\quad 1\hspace{1em}2\hspace{1em}3\hspace{1em}4\hspace{1em}5\hspace{1em}6\hspace{1em}7\quad Strongly Agree
\vspace{1em}

Q5: I can clearly identify areas of my writing that need improvement when using the AI tool.

Strongly Disagree\quad 1\hspace{1em}2\hspace{1em}3\hspace{1em}4\hspace{1em}5\hspace{1em}6\hspace{1em}7\quad Strongly Agree
\vspace{1em}

Q6: After writing, I reviewed the narrative to assess how well it communicated the scientific content.

Strongly Disagree\quad 1\hspace{1em}2\hspace{1em}3\hspace{1em}4\hspace{1em}5\hspace{1em}6\hspace{1em}7\quad Strongly Agree
\vspace{1em}

Q7: I am able to adjust my writing strategies during the editing process.

Strongly Disagree\quad 1\hspace{1em}2\hspace{1em}3\hspace{1em}4\hspace{1em}5\hspace{1em}6\hspace{1em}7\quad Strongly Agree
\vspace{2em}

\textbf{Part 2: Control}
(Control: )

Q8: I felt in control of the writing process while interacting with the system.

Strongly Disagree\quad 1\hspace{1em}2\hspace{1em}3\hspace{1em}4\hspace{1em}5\hspace{1em}6\hspace{1em}7\quad Strongly Agree
\vspace{1em}

Q9: I was able to override or ignore the system’s suggestions when I thought it was necessary.

Strongly Disagree\quad 1\hspace{1em}2\hspace{1em}3\hspace{1em}4\hspace{1em}5\hspace{1em}6\hspace{1em}7\quad Strongly Agree
\vspace{1em}

Q10: I determined the direction and flow of the science narrative, not the system.

Strongly Disagree\quad 1\hspace{1em}2\hspace{1em}3\hspace{1em}4\hspace{1em}5\hspace{1em}6\hspace{1em}7\quad Strongly Agree
\vspace{2em}

\textbf{Part 3: Autonomy}
(Autonomy: )

Q11: I felt free to make my own choices during the co-writing process with the system.

Strongly Disagree\quad 1\hspace{1em}2\hspace{1em}3\hspace{1em}4\hspace{1em}5\hspace{1em}6\hspace{1em}7\quad Strongly Agree
\vspace{1em}

Q12: The system supported my ability to express my own ideas in the narrative.

Strongly Disagree\quad 1\hspace{1em}2\hspace{1em}3\hspace{1em}4\hspace{1em}5\hspace{1em}6\hspace{1em}7\quad Strongly Agree
\vspace{1em}

Q13: I did not feel pressured to accept the system’s suggestions.

Strongly Disagree\quad 1\hspace{1em}2\hspace{1em}3\hspace{1em}4\hspace{1em}5\hspace{1em}6\hspace{1em}7\quad Strongly Agree












\subsection{Participant Demographic Information}\label{Participants}

\begin{table*}[th]
\label{Demographic}
\centering
\footnotesize
\begin{tabular}{rrlllll}
\toprule
   \textbf{ID}  & \textbf{Age} & \textbf{Gender} & \textbf{Education} 
   & \textbf{AI Writing Use} & \textbf{Writing Confidence} & \textbf{Occupation} \\
\midrule
    1 & 26 & Male   & Postgraduate             & Occasionally & Confident & (a) \\
    2 & 27 & Male   & Postgraduate             & Daily        & Confident & (a), (b), (c), (d) \\
    3 & 26 & Male   & Postgraduate             & Daily        & Confident & (b), (d) \\
    4 & 25 & Female & Postgraduate             & Daily        & Confident & (a), (b), (c) \\
    5 & 24 & Male   & Postgraduate             & Daily        & Confident & (a) \\
    6 & 28 & Female & Postgraduate             & Weekly       & Neutral   & (a) \\
    8 & 28 & Male   & Postgraduate             & Occasionally & Neutral   & (a) \\
    7 & 29 & Female & Higher than postgraduate & Daily        & Confident & (a), (b) \\
    9 & 31 & Male   & Postgraduate             & Weekly       & Neutral   & (a) \\
   10 & 24 & Female & Postgraduate             & Occasionally & Confident & (a), (c) \\
   11 & 29 & Female & Postgraduate             & Weekly       & Neutral   & (a) \\
   12 & 26 & Male   & Postgraduate             & Weekly       & Neutral   & (a) \\
   14 & 27 & Male   & Postgraduate             & Daily        & confident & (a), (b) \\
   15 & 24 & Female & Postgraduate             & Weekly       & Neutral   & (a) \\
   16 & 30 & Male   & Postgraduate             & Weekly       & Neutral   & (a) \\
\bottomrule
\end{tabular}

\vspace{0.5em}
\begin{flushleft}
\textbf{\textit{Occupation: }}
(a) PhD Student / Postdoctoral Researcher / University Faculty / Researcher;\\
(b) Science Journalist / Media Producer;\\
(c) Educator / Teacher;\\
(d) Online Science Content Creator (e.g., YouTube, Blog, TikTok, etc.)
\end{flushleft}
\end{table*}

\clearpage
\subsection{User Study Results}\label{Interaction}
1. Visualization of interaction behaviors from 16 participants across two revision directions:

\begin{figure*}[h]
  \centering
  \includegraphics[width=0.9\linewidth]{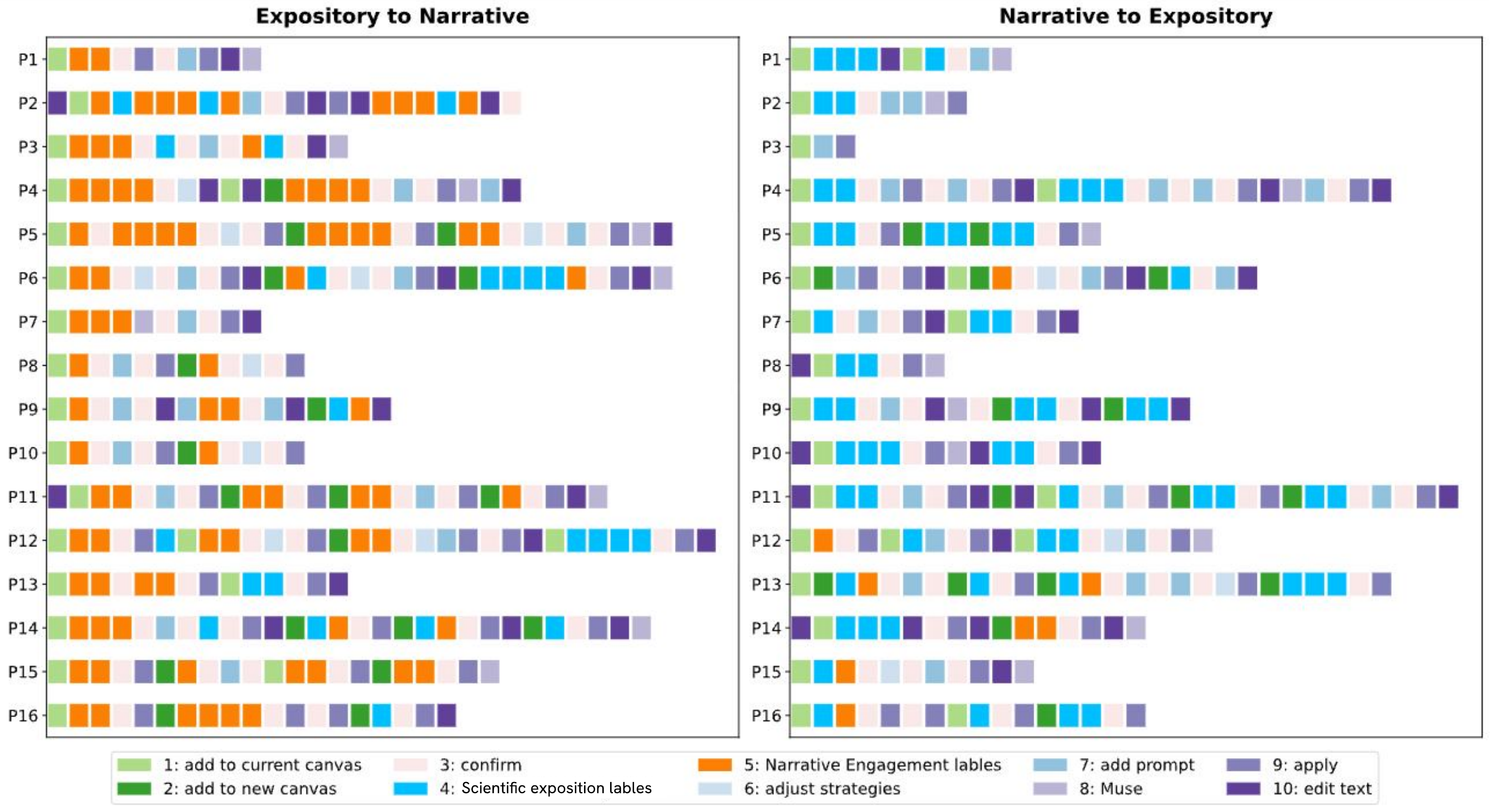}
  \caption{Visualization of interaction behaviors from 16 participants across two revision directions.}
  \label{fig:User_Behavior}
  \Description{Interaction timelines for 16 participants across two revision directions: expository to narrative on the left and narrative to expository on the right. Each row represents one participant, and each colored block represents a specific interaction action in sequence, such as adding text to the canvas, adding to a new canvas, confirming a version, using scientific exposition labels, using narrative engagement labels, adjusting strategies, adding prompts, using Muse, applying revisions, or editing text directly. The left panel generally shows more frequent use of narrative engagement labels, while the right panel shows more frequent use of scientific exposition labels, reflecting adaptation to the revision goal. Across both panels, participants combine multiple actions in varied orders, indicating diverse but structured iterative workflows rather than a single fixed interaction pattern.}
\end{figure*}

\begin{figure*}[h]
  \centering
  \includegraphics[width=1\linewidth]{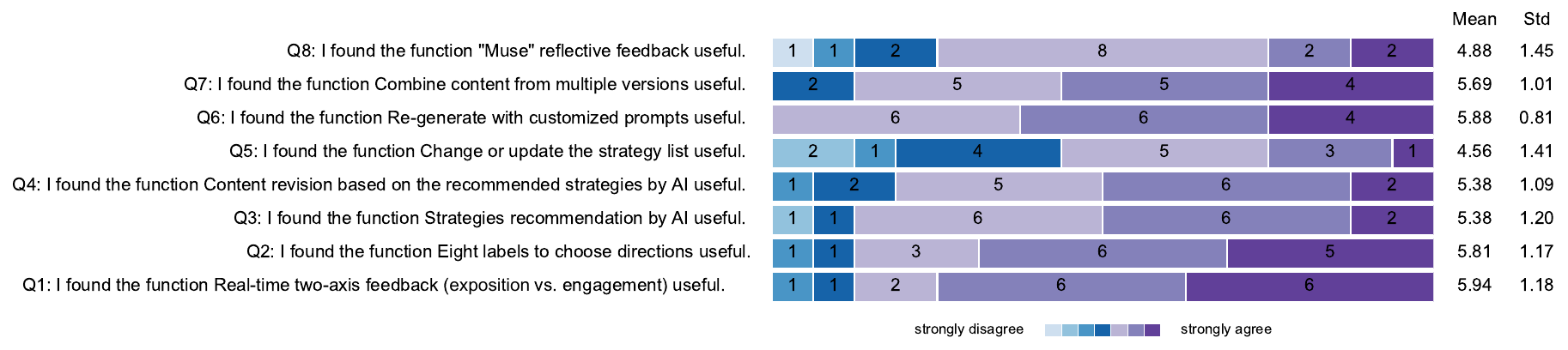}
  \caption{Functional Evaluation of \system{}.}
  \label{fig:Function}
  \Description{Functional evaluation of eight Spatial Balancing features shown as stacked horizontal bar charts on a seven-point agreement scale from strongly disagree to strongly agree. The evaluated features are real-time two-axis feedback, eight rhetorical labels for choosing directions, strategy recommendation by AI, content revision based on recommended strategies, changing or updating the strategy list, regenerating with customized prompts, combining content from multiple versions, and Muse reflective feedback. Most responses cluster toward agreement, indicating generally positive evaluations across features. The highest mean ratings are for real-time two-axis feedback and the eight labels, while changing or updating the strategy list receives the lowest mean rating among the eight items. Means and standard deviations are listed on the right.}
\end{figure*}

\clearpage
\subsection{Prompts}\label{Prompts}
\subsubsection{Recommender}
\ 
\newline
\indent The blue word will be replaced by input information.

\begin{tcolorbox}[colback=cyan!10, colframe=darkblue, coltitle=newgreen]

\small
\begin{Verbatim}[commandchars=\\\{\}]
\textcolor{purple}{# Base prompt}
You are an expert in science communication narrative text revision and strategy recommendation. 
Your task is to analyze the given text and recommend effective strategies to improve it. 

\textcolor{purple}{# Order prompt}
Step 1: Analyze the Text. 
Position: Identify where the selected text \textcolor{blue}{\{text\}} appears in the \textcolor{blue}{\{overall_content\}}. 
Granularity: Determine whether the text consists of sentences, paragraphs, or a complete document. 
Core Message: Extract the key ideas that must be preserved and effectively conveyed in text. 

Step 2: Select Strategies Review the available strategy list \textcolor{blue}{\{strategy_info\}}, including their 
definitions, examples, and usage instructions. Choose a set of strategies that align with the 
text's characteristics and modification goals. Ensure the selected strategies are compatible 
when combined. Consider multiple ways to apply the strategies for improvement. 
Only choose strategies mentioned above, and use them appropriately. 
Provide \textcolor{blue}{\{generated_number\}} different versions, each using distinct or complementary strategy sets. 
These different versions should use different strategies, preferably with varied combinations of 
strategies. 

Step 3: Output the Strategy List Return the strategy selection in JSON format with multiple versions: 
\{ 
"Version1": [ "Strategy_A", "Strategy_H", "Strategy_J", "Strategy_B"],
"Version2": [ "Strategy_F",..., "Strategy_E"],
...,
"Version_number": [ "Strategy_G", "Strategy_M",..., "Strategy_C",...,"Strategy_D"] 
\} 
Do not include any extra commentary or explanation outside the JSON. 
Let's think step by step.
\end{Verbatim}
\end{tcolorbox}

\newpage
\subsubsection{Generator}
\ 
\newline
\indent The blue word will be replaced by input information.

\begin{tcolorbox}[title={Generate new text based on user selected goals},colback=cyan!10, colframe=darkblue,coltitle=newgreen]

\small
\begin{Verbatim}[commandchars=\\\{\}]
\textcolor{purple}{# Order prompt}
You are an expert in science communication narrative strategy. Your task is to revise the 
given text using the recommended strategies and provide a concise overview of how the 
strategies were applied. 

Step 1: Review the Strategy List 
- Read the strategy list \textcolor{blue}{\{strategy_info\}}, including each strategy's definition and 
how it is typically used.

Step 2: Apply all the Strategies mentioned in the strategy list to the Text: \textcolor{blue}{\{text\}}.
Even if the original text already contains elements that align with the strategy, enhance it further 
based on how the strategy should be applied. 
Also, consider the position of the given text in the whole context \textcolor{blue}{\{overall_content\}}. 
Make the changed text coherent with the context. 

Step 3: Summarize the Application 
- Summarize how each selected strategy was applied. 
- Keep the summary concise and short to indicate what specific changes have been made using 
separate strategies. 

Step 4: Do not omit or alter any important information from the original text, but ensure that the 
generated text is distinct from the original. 

Step 5: If the content is primarily narrative in nature, supplement it with scientifically grounded 
explanations, relevant data, or reliable sources to enhance credibility and depth.  

Step 6: Output the Result Return a JSON with the following structure: 
\{
"strategies": ["Strategy_A", ..., "Strategy_B", "Strategy_C", "Strategy_D"], 
"summary": "Summarize how each strategy was applied and what specific changes were made to the content 
\qquad \qquad \quad based on each strategy. Example: Changed 'Photosynthesis is the process plants use to 
\qquad \qquad \quad make food.' to 'What if plants could teach us how to turn sunlight into fuel?
\qquad \qquad \quad Focus only on the changes from the previous version.'", 
"newText": "Modified version of the text. Even if the original text already contains elements that 
\qquad \qquad \quad align with the strategy, enhance it further based on how the strategy should be applied."
\}

Do not include any extra commentary or explanation outside the JSON. 
Let's think step and step.

\end{Verbatim}
\end{tcolorbox}

\newpage
\subsubsection{Scorer}
\ 
\newline
\indent The blue word will be replaced by input information.

\begin{tcolorbox}[colback=cyan!10, colframe=darkblue,coltitle=newgreen]

\small
\begin{Verbatim}[commandchars=\\\{\}]
\textcolor{purple}{# Base prompt}
You are an engaging audience for science communication. 
Given a narrative, evaluate it on two dimensions: (1) Narrative Engagement and (2) Scientific Exposition. 
using the detailed scoring rubrics below. 
Provide a numerical score from 0 to 100 for each dimension, along with a brief explanation justifying 
your rating. 

Dimension 1: 
Narrative Engagement: Evaluate how effectively the narrative captures attention, evokes emotion, 
sparks curiosity, and maintains reader engagement. 
Scoring Rubric: 
0-20: Extremely boring and dry, no storytelling elements,
21-40: Barely engaging, logical but lacks emotion or creativity,
41-60: Moderately engaging, uses some analogies or description but still feels academic,
61-80: Quite engaging, includes storytelling techniques and relatable examples,
81-100: Highly immersive, vivid storytelling with strong emotional or narrative appeal.

Dimension 2: Scientific Exposition: Assess how well the narrative explains scientific concepts with 
clarity, 
correctness, and alignment with established knowledge. 
Scoring Rubric: 
0-20: Highly inaccurate or pseudoscientific, major factual errors,
21-40: Misleading or speculative, lacks clarity or evidence,
41-60: Mostly accurate but vague or oversimplified,
61-80: Generally accurate, minor imprecision, lacks citations,
81-100: Highly accurate, precise, and well-aligned with scientific consensus. 

\textcolor{purple}{# Order prompt}
This is the original text: \textcolor{blue}{\{text\}} and its score \textcolor{blue}{\{currentScore\}}. Please use this as a reference. 
Compare the current version with the original one in terms of scientific exposition and narrative 
engagement, and assess whether it performs better or worse than the previous version. 
Compared to the previous version's scores, assign a score difference within a reasonable range.

\end{Verbatim}
\end{tcolorbox}

%% file: figs/survey.tex
\definecolor{Story}{HTML}{FC8002}
\definecolor{Expository}{HTML}{4995C6}

\definecolor{Short}{HTML}{FC8002}
\definecolor{Medium}{HTML}{EE4431}
\definecolor{Long}{HTML}{4995C6}


\begin{tikzpicture}

        \begin{scope}
        \begin{axis}[
            width=5cm,
            height=5cm,
            title={By Narrative Style},
            xlabel={Narrative Engagement},
            ylabel={Scientific Expostion},
            grid=both,
            grid style={dashed, opacity=0.7},
            xmin=20, xmax=70,
            ymin=20, ymax=80,
            xtick={20,30,40,50,60,70},
            ytick={20,30,40,50,60,70,80},
            legend style={
                draw=none,
                at={(0.4,1.4)},
                anchor=north,
                legend columns=3
            }
        ]
        \addplot[
            scatter,
            only marks,
            mark=*,
            mark size=2,
            opacity=0.8,
            draw=none,
            line width=0.5pt,
            scatter src=explicit symbolic,
            scatter/classes={
                Expository={mark=*, fill=Expository},
                Medium={mark=*, fill=Medium},
                Story={mark=*, fill=Story}
            }
        ] table[
            x=Average_NE,
            y=Average_SA,
            meta=Tag2,
            col sep=comma
        ] {data.csv};

        \addlegendimage{only marks, mark=*, mark size=2, fill=Expository}
        \addlegendentry{Expository}
        \addlegendimage{only marks, mark=*, mark size=2, fill=Medium}
        \addlegendentry{Infotainment }
        \addlegendimage{only marks, mark=*, mark size=2, fill=Story}
        \addlegendentry{Story}
        \end{axis}
        \end{scope}

        \begin{scope}[shift={(4cm,0)}]
        \begin{axis}[
            width=5cm,
            height=5cm,
            title={By Length},
            xlabel={Narrative Engagement},
            yticklabels={},
            grid=both,
            grid style={dashed, opacity=0.7},
            xmin=20, xmax=70,
            ymin=20, ymax=80,
            xtick={20,30,40,50,60,70},
            ytick={20,30,40,50,60,70,80},
            legend style={
                draw=none,
                at={(0.52,1.4)},
                anchor=north,
                legend columns=3
            }
        ]
        \addplot[
            scatter,
            only marks,
            mark=*,
            mark size=2,
            opacity=0.8,
            draw=none,
            line width=0.5pt,
            scatter src=explicit symbolic,
            scatter/classes={
                Long={mark=*, fill=Long},
                Medium={mark=*, fill=Medium},
                Short={mark=*, fill=Short}
            }
        ] table[
            x=Average_NE,
            y=Average_SA,
            meta=Tag1,
            col sep=comma
        ] {data.csv};

        \addlegendimage{only marks, mark=*, mark size=2, fill=Long}
        \addlegendentry{Long}
        \addlegendimage{only marks, mark=*, mark size=2, fill=Medium}
        \addlegendentry{Medium}
        \addlegendimage{only marks, mark=*, mark size=2, fill=Short}
        \addlegendentry{Short}
        \end{axis}
        \end{scope}

    \end{tikzpicture}